\documentclass[conference,onecolumn]{IEEEtran}
\usepackage{amsmath,amssymb,amsfonts,mathtools}
\usepackage{array,multirow,diagbox}
\usepackage{hyperref}
\hypersetup{hidelinks=true}
\def\BibTeX{{\rm B\kern-.05em{\sc i\kern-.025em b}\kern-.08em
		T\kern-.1667em\lower.7ex\hbox{E}\kern-.125emX}}
\usepackage{balance}
\usepackage[nospace]{cite}

\newcommand{\E}[2]{\ensuremath{{E}_{#2}\left[#1\right]}}
\usepackage{romannum}

\usepackage{acronym}
\acrodef{fmcw}[FMCW]{frequency modulated continuous wave}
\acrodef{mmic}[MMIC]{monolithic microwave integrated circuit}
\acrodef{adc}[ADC]{analog-to-digital converter}
\acrodef{dac}[DAC]{digital-to-analog converter}
\acrodef{sfdr}[SFDR]{spurious-free dynamic range}
\acrodef{sndr}[SNDR]{signal to noise and distortion ratio}
\acrodef{seir}[SEIR]{stimulus error identification and removal}
\acrodef{sh}[S\&H]{sample and hold}
\acrodef{hee}[HEC]{homogeneity enforced calibration}
\acrodef{hec}[HEC]{homogeneity enforced calibration}
\acrodef{mdac}[MDAC]{multiplying DAC}
\acrodef{enob}[ENOB]{effective number of Bits}
\acrodef{inl}[INL]{integral nonlinearity}
\acrodef{lms}[LMS]{least mean squares}
\acrodef{pn}[PN]{phase noise}
\acrodef{dpn}[DPN]{decorrelated phase noise}
\acrodef{if}[IF]{intermediate frequency}
\acrodef{dut}[DUT]{device under test}
\acrodef{tsg}[TSG]{test signal generator}
\acrodef{mmse}[MMSE]{minimum mean sqared error}
\acrodef{mse}[MSE]{mean squared error}
\acrodef{sa}[SA]{signal of the application}
\acrodef{sgd}[SGD]{stochastic gradient descent}
\acrodef{pdf}[PDF]{probability density function}
\acrodef{als}[ALS]{approximate least squares}
\acrodef{pvt}[PVT]{process, voltage, and temperature}
\acrodef{xhec}[BL-HEC]{bi-linear homogeneity enforced calibration}
\acrodef{rls}[RLS]{recursive least squares}
\acrodef{snr}[SNR]{signal to noise ratio}
\acrodef{afe}[AFE]{analog front-end}

\begin{document}
\title{Bi-Linear Homogeneity Enforced Calibration for Pipelined ADCs}
\author{Matthias Wagner$^1$, Oliver Lang$^1$, Esmaeil Kavousi Ghafi$^1$,Arianit Preniqi$^{2,3,*}$, Andreas Schwarz$^2$ and\\ Mario Huemer$^1$ \vspace{0.1cm} \\
	\normalsize $^1$ Institute of Signal Processing, Johannes Kepler University, Linz, Austria\\
	$^2$Infineon Technologies Linz GmbH \& Co. KG, Austria.\\
	$^3$Institute for Communication Engineering and RF-Systems, Johannes Kepler University, Linz, Austria. }
\markboth{}%
{}

\maketitle
\makeatletter{\renewcommand*{\@makefnmark}{}
	\footnotetext{$^*$His financial support by the Austrian Federal Ministry of Labour and Economy, the  National Foundation for Research, Technology and Development and the Christian Doppler Research Association is gratefully acknowledged.}\makeatother}
	\begin{abstract}
		Pipelined \acp{adc} are key enablers in many state-of-the-art signal processing systems with high sampling rates. In addition to high sampling rates, such systems often demand a high linearity. To meet these challenging linearity requirements, \ac{adc} calibration techniques were heavily investigated throughout the past decades.\\
		One limitation in \ac{adc} calibration is the need for a precisely known test signal. In our previous work, we proposed the \ac{hec} approach, which circumvents this need by consecutively feeding a test signal and a scaled version of it into the \ac{adc}. The calibration itself is performed using only the corresponding output samples, such that the test signal can remain unknown. On the downside, the \ac{hec} approach requires to accurately scale the test signal, impeding an on-chip implementation.\\
		In this work, we provide a thorough analysis of the \ac{hec} approach, including limitations such as the effects of an inaccurately scaled test signal. Furthermore, the \ac{xhec} approach is introduced and suggested to account for an inaccurate scaling and, therefore, to facilitate an on-chip implementation. In addition, a comprehensive stability analysis of the \ac{xhec} approach is carried out. Finally, we verify our concept with behavioral Matlab simulations and measurements conducted on 24 integrated \acp{adc}.
	\end{abstract}
	

	\maketitle
	
	\acresetall
	\section{Introduction}
	\label{se:Introduction}
	High-resolution, high-speed pipelined \acp{adc} are key building blocks in many of today's signal processing systems. Although the time-interleaved architecture of pipelined \acp{adc} enables the high sampling rates, it also introduces additional error sources. Besides random errors, i.e., thermal or quantization noise, the \ac{adc}'s performance is limited by systematic errors of its analog building blocks \cite{14_ADC_HighSpeedTimeInterleavedADCs}. To obtain a high \ac{adc} performance, calibration techniques became an essential part of modern \acp{adc} and were heavily investigated in recent years \cite{19_ADC_CalibrationAndDynamicMatchinInDataConverters1,20_ADC_CalibrationAndDynamicMatchinInDataConverters2}. Concerning power consumption and scalability, calibration techniques are preferably carried out in the digital domain \cite{12_ADC_ASurveyOnDigitalBackgroundCalibrationOfADCs}. Typically, calibration consists of two parts, that are the identification of non-idealities, and their correction. On account of the identification's working principle, calibration techniques may be grouped into histogram-, correlation-, or equalization-based approaches. In general, these techniques involve either long test times \cite{4_ADC_DigitalBackgroundCalibrationWithHistogramOfDecisionPointsInPipelinedADCs,5_ADC_ADigitalBackgroundCalibrationSchemeForPipelinedADCsUsingMultipleCorrelationEstimation,7_ADC_StatisticsBasedDigitalBackgroundCalibrationOfResidueAmplifierNonlinearityInPipelinedADCs}, additional hardware components \cite{3_ADC_AnAdaptiveDigitalBackgroundCalibrationTechniqueUsingVariableStepSizeLMSForPipelinedADC,1_ADC_LeasMeanSquareAdaptiveDigitalBackgroundCalibrationOfPipelinedAnalogToDigitalConverters}, or an exceedingly linear \ac{tsg} \cite{21_ADC_AlgorithmForDramaticallyImprovedEfficiencyInADCLinearityTest,22_ADC_LinearityTestingIssuesOfAnalogToDigitalConverters}. The latter impedes an on-chip implementation for extremely linear \acp{adc}, as a \ac{tsg} with higher linearity than the \ac{adc} is required. Therefore, much effort was devoted to relax these linearity requirements \cite{13_ADC_ADCTestMethodsUsingAnImpureStimulus}. In \cite{15_ADC_AccurateTestingOfAnalogTotDigitalConvertersUsingLowLinearitySignalsWithStimulusErrorIdentification}, the need for an exceedingly linear \ac{tsg} is omitted by injecting the test signal twice into the \ac{adc}, whereas for the second time, a constant voltage offset is applied. Therewith, the \ac{tsg}'s nonlinearity can be identified and compensated. In literature, this is referred to as \ac{seir} approach. Improvements  of the \ac{seir} approach, such as enhanced robustness concerning reference voltage stationarity, or a reduced hardware overhead, were investigated in \cite{18_ADC_HighPrecisionADCTestingWithRelaxedReferenceVoltageStationarity,16_ADC_ARobustAlgorithmToIdentifyTheTestStimulusInHistogramBasedADCTesting}, respectively. In \cite{11_ADC_USERSMILEUltrafastStimulusErrorRemovalAndSegmentedModelIdentificationOfLinearityErrorsForADCBuiltInSelfTest}, the principle of the \ac{seir} approach is combined with a segmented nonlinearity model, which leads to a significantly reduced test time. 
	\\
	To analyze the \ac{seir}-based approaches, the \ac{adc} transition function may be represented as a nonlinear function $f(x): \mathbb{R}\rightarrow\mathbb{R}$. Obviously, this function violates the linearity conditions, by means that either
	\begin{align}
		f(x+y) & =  f(x) + f(y) &&\ldots\text{ additivity}
		\label{eq_add}\\
		\intertext{and/or}
		f(\alpha x) & =  \alpha f(x)  &&\ldots\text{ homogeneity}
		\label{eq_hom}
	\end{align}
	with $x$, $y$ as the system inputs and $\alpha\in\mathbb{R}$ a constant, do not hold. For the \ac{seir}-based approaches, $x$ and $y$ in (\ref{eq_add}) represent the test signal and the applied voltage offset, respectively. In this sense, they utilize the lack of additivity to identify and correct the \ac{tsg}'s nonlinearity.  \\
	In \cite{1_HEC}, we proposed the \ac{hec} approach, which is based on the homogeneity condition in (\ref{eq_hom}). As with the \ac{seir}-based approaches, the test signal is injected twice into the \ac{adc}. However, at the second time, it is scaled by a constant factor $\alpha$ in the analog domain. By shifting the focus from additivity to homogeneity, the \ac{hec} approach omits the necessity of a highly constant voltage shift. On the downside, a very precisely known scaling factor $\alpha$ is mandatory. The implementation of such an accurate scaling factor is a challenging task because of \ac{pvt} variations. Thus, an on-chip utilization of the \ac{hec} approach is limited in real-world applications. \\
	In this work, we propose the \ac{xhec} approach, which accounts for an inaccurate scaling factor $\alpha$. Note that, the \ac{xhec} approach maintains the fast convergence times of the original \ac{hec} approach while it restores the high \ac{sfdr} improvement despite an inaccurately scaled test signal. Furthermore, the \ac{xhec} approach does not require any additional analog hardware components compared to its original version.
	The key contributions of this work are summarized below: 
	\begin{itemize}
		\item Further analysis of the \ac{hec} approach is provided, highlighting limitations such as an inaccurate scaling factor.
		\item The \ac{xhec} approach is introduced to address the inaccurate scaling factor, which is crucial for the practical relevance of the overall concept. 
		\item Two novel algorithms -- the \ac{xhec} Wiener filter and the \ac{xhec} \ac{sgd} approach -- are derived.
		\item The findings are supported by simulations and by measurements conducted on 24 \acp{adc} in radar sensors.
	\end{itemize}
	The rest of this paper is organized as follows. In Section\,\ref{se:systemModel}, the output signal of a non-ideal \ac{adc} is briefly discussed. An equivalent representation of the output signal is derived, which motivates the presented \ac{adc} post-correction model. The \ac{hec} approach is shortly reviewed in Section\,\ref{se:HEC}, followed by a novel, in-depth analysis of its limitations such as the impact of a scaling factor mismatch. In Section\,\ref{se:XHEC}, two novel algorithms, which can cope with the scaling factor mismatch, are introduced. All findings are numerically verified in the first part of Section\,\ref{se:Results}. The second part of Section\,\ref{se:Results} shows the effectiveness of the proposed concept based on measurement results obtained from 24 \acp{adc} employed in $77\,\text{GHz}$ radar monolithic microwave integrated circuits. Finally, Section\,\ref{se:Con} concludes the work.
	\section{SYSTEM MODEL}
	\label{se:systemModel}
	In this section, the basic architecture of a pipelined \ac{adc} is shortly revised. Furthermore, the interpretation of the non-ideal \ac{adc} model, introduced in \cite{1_HEC}, is extended to support the stability analysis provided in Section\,\ref{se:XHEC}. Note that additional \ac{adc} non-idealities are considered in \cite{2_HEC,ADC_DigitalBackgroundCalibrationForMemoryEffects}, i.e., nonlinear stage amplifiers and memory effects, respectively. These effects are not addressed in this work for readability reasons, but can easily be included, as discussed in \cite{2_HEC,ADC_DigitalBackgroundCalibrationForMemoryEffects}. Ultimately, the post-correction model employed in this work is derived in detail. 
	\subsection{PIPELINED ADC MODEL}
	\begin{figure}[!t]
		\centering
		\includegraphics{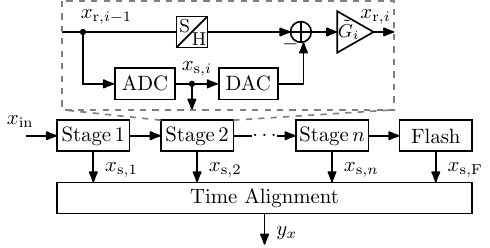}
		\caption{Schematic of the pipelined \ac{adc} architecture including internal key building blocks of the stages \cite{1_HEC}.}
		\label{fig:PipelinedADCSchematic}
	\end{figure}
	The main building blocks of a pipelined \ac{adc} are equally structured stages, as shown in Fig.~\ref{fig:PipelinedADCSchematic}. In the $i$th stage, the input signal $x_{\text{r},i-1}$ is digitized with a stage \ac{adc}, resulting in the stage output signal $x_{\text{s},i}$. Note that, for the first stage $x_{\text{r},0} = x_\text{in}$, with $x_\text{in}$ as the \ac{adc} input signal. The stage output signal is transformed back to the analog domain with a stage \ac{dac} and subtracted from the stage input signal $x_{\text{r},i-1}$, which is stored in a \ac{sh} block. The difference signal is amplified by the stage gain $\tilde{G}_i$ and subsequently fed into the next stage. In literature, this amplified difference signal is typically referred to as the residue signal. The last stage only comprises a flash \ac{adc}. Finally, the stage output signals $x_{\text{s},i}$ are used to compute the overall \ac{adc} output $y_x$. Due to the \ac{sh} blocks, the outputs of consecutive stages are time delayed. Thus, the time alignment in Fig.~\ref{fig:PipelinedADCSchematic} is a signature building block of pipelined \acp{adc}. As derived in \cite{1_HEC}, by neglecting the time alignment for the sake of better readability, the overall system output may be written as
	\begin{equation}
		\begin{split}
			y_x = &\text{\space} x_\text{in}-e_{\text{s,F}}^\text{q}\prod_{j = 1}^{n}\frac{1}{G_j}+\sum_{i=1}^{n}\left(e_{\text{s},i}^\text{q}\zeta_i-\left(1+\zeta_i\right)e_{\text{s},i}^\text{DA}\right)\prod_{j = 1}^{i-1}\frac{1}{G_j},
		\end{split}
		\label{eq:outputSignal3}
	\end{equation}
	with $e_{\text{s},i}^\text{q}$ as the digitization error of the $i$th stage, $e_{\text{s,F}}^\text{q}$ as the digitization error of the final flash \ac{adc} stage, $e_{\text{s},i}^\text{DA}(x_{\text{s},i})$ representing the \ac{dac} mismatches of the $i$th stage, and the gain mismatch $\zeta_i$ between the ideal stage gain $G_i$ and the true stage gain $\tilde{G}_i$.
	There, the \ac{adc} output signal is expressed as the ideal output signal $x_\text{in}-e_{\text{s,F}}^\text{q}\prod_{j = 1}^{n}\frac{1}{G_j}$ and a summation of all stage non-idealities. These stage non-idealities are weighted by the product of the preceding stage gain reciprocals. Thus, errors in the first few stages primarily affect the overall \ac{adc} nonlinearity, while errors in the last few stages may be neglected.  Consequently, calibrating the most significant stages can yield sufficient calibration performances in many applications \cite{2_ADC_EqualizationBasedDigitalBackgroundCalibrationTechniqueForPipelinedADCs,9_ADC_BlackBoxCalibrationForADCsWithHardNonlinearErrorsUsingANovelINLBasedAdditiveCode,3_ADC_AnAdaptiveDigitalBackgroundCalibrationTechniqueUsingVariableStepSizeLMSForPipelinedADC}. Furthermore, the effects of the considered non-idealities are schematically illustrated in \cite[Fig.~2 a) and b)]{2_HEC}.
	\subsection{POST-CORRECTION MODEL}
	\label{sec:postcorrmodel}
	To develop a post-correction model for the considered non-idealities, the pipelined \ac{adc} model is rearranged in the following. As shown with the summation term in (\ref{eq:outputSignal3}), the non-idealities consist of the stage digitization error $e_{\text{s},i}^\text{q}$, the relative gain mismatch $\zeta_i$ and the \ac{dac} mismatches $e_{\text{s},i}^\text{DA}$. The stage digitization error as well as the \ac{dac} mismatches are functions of the stage input, implying that different correction values are required for different input values. In order to deduce a meaningful post-correction model from the \ac{adc} model, these different correction values are put together in a parameter vector. Therefore, $x_{\text{s},i}$ is rewritten in vector notation as 
	\begin{equation}
		x_{\text{s},i} = \mathbf{w}_i^T\mathbf{d}_i,
		\label{eq:xstageVec}
	\end{equation}
	with $\mathbf{d}_i\in \mathbb{R}^{p_i}$ as a vector containing the digital values of the $i$th stage \ac{adc}, $p_i$ as the number of quantization levels in the $i$th stage, the selection subvector $\mathbf{w}_i\in\left\{0,1\right\}^{p_i}$ and $(.)^T$ denoting transposition. The elements of $\mathbf{w}_i$ are 
	\begin{equation*}
		\left[\mathbf{w}_i\right]_j = \begin{cases}
			1 & \text{if $x_{\text{r},i-1}\leq  \left[\mathbf{v}_i\right]_{j}\land j = 1$,}\\
			1 & \text{if $\left[\mathbf{v}_i\right]_{j-1} < x_{\text{r},i-1}\leq  \left[\mathbf{v}_i\right]_{j}\land j = 2\ldots p_{i}-1$,}\\
			1 & \text{if $\left[\mathbf{v}_i\right]_{j-1} < x_{\text{r},i-1}\land j = p_{i}$}\\
			0 & \text{otherwise}.
		\end{cases}   
		\label{eq:selectionVector}    
	\end{equation*}
	where  $\mathbf{v}_i\in \mathbb{R}^{p_i-1}$ is a vector containing the \ac{adc} threshold voltages $\left[\mathbf{v}_i\right]_{j}$ of the $i$th stage. With $x_{\text{s},i} = \frac{1}{v_\text{ref}}\left(x_{\text{r},i-1}-e_{\text{s},i}^\text{q}(x_{\text{r},i-1})\right)$, $v_\text{ref} = 1\,\text{V}$, and (\ref{eq:xstageVec}), the stage digitization error can be derived as 
	\begin{equation}
		e_{\text{s},i}^\text{q} = x_{\text{r},i-1} -\mathbf{w}_i^T\mathbf{d}_i.
		\label{eq:stageOutput2}
	\end{equation}
	Similarly, writing the \ac{dac} mismatches in vector notation yields
	\begin{equation}
		e_{\text{s},i}^\text{DA} = \mathbf{w}_i^T \mathbf{e}_i^\text{DA},
		\label{eq:dacerror}
	\end{equation}
	with the same selection subvector $\mathbf{w}_i$ and an error vector $\mathbf{e}_i^\text{DA}\in\mathbb{R}^{p_i}$. Finally, (\ref{eq:outputSignal3}) can be written in the required form such that all errors occur in a vector. In the following, this derivation is exemplarily carried out for $n=2$ stages. In that case, (\ref{eq:outputSignal3}) becomes
	\begin{equation}
		\begin{split}
			y_x =&\text{\space} x_\text{in}-e_{\text{s,F}}^\text{q}\frac{1}{G_1G_2}+e_{\text{s},1}^\text{q}\zeta_1-e_{\text{s},1}^\text{DA}\left(1+\zeta_1\right)+\frac{1}{G_1}\left(e_{\text{s},2}^\text{q}\zeta_2-e_{\text{s},2}^\text{DA}\left(1+\zeta_2\right)\right).
		\end{split}
		\label{eq:outputSignal2st}
	\end{equation} 
	Given $x_{\text{r},i} = \tilde{G}_i\left(e_{\text{s},i}^\text{q}(x_{\text{r},i-1})-e_{\text{s},i}^\text{DA}(x_{\text{s},i})\right)$, and (\ref{eq:stageOutput2}), the digitization error of the first and second stage may be written as 
	\begin{align}
		e_{\text{s},1}^\text{q} &= x_\text{in}-\mathbf{w}_1^T\mathbf{d}_1\label{eq:stageOutput11}\\
		e_{\text{s},2}^\text{q} &= \tilde{G}_1\left( x_\text{in}-\mathbf{w}_1^T\left(\mathbf{d}_1+\mathbf{e}_{1}^\text{DA}\right)\right)-\mathbf{w}_2^T\mathbf{d}_2,\label{eq:stageOutput12}
	\end{align}
	respectively. With $\tilde{G}_i = G_i\left(1+\zeta_i\right)$, (\ref{eq:dacerror}), (\ref{eq:stageOutput11}), and (\ref{eq:stageOutput12}) the \ac{adc} output in (\ref{eq:outputSignal2st}) becomes
	\begin{equation*}
		\begin{split}
			y_x =&\text{\space}x_\text{in}\left[1+\zeta_1+\left(1+\zeta_1\right)\zeta_2\right]-e_{\text{s,F}}^\text{q}\frac{1}{G_1G_2}-\mathbf{w}_1^T\left[\left(\zeta_1+\left(1+\zeta_1\right)\zeta_2\right)\mathbf{d}_1+\left(1+\zeta_1\right)\left(1+\zeta_2\right)\mathbf{e}_{1}^\text{DA}\right]-\\
			&\text{\space}\mathbf{w}_2^T\frac{1}{G1}\left[\zeta_2\mathbf{d}_2+\left(1+\zeta_2\right)\mathbf{e}_{2}^\text{DA}\right].
		\end{split}
		\label{eq:outputSignal4}
	\end{equation*}
	Substituting 
	\begin{align*}
		\beta &= 1+\zeta_1+\left(1+\zeta_1\right)\zeta_2\in\mathbb{R},\\
		\boldsymbol{\varphi}_0 &= \begin{bmatrix}
			\left(\beta-1\right)\mathbf{d}_1+\left(1+\zeta_1\right)\left(1+\zeta_2\right)\mathbf{e}_{1}^\text{DA}\\
			\frac{1}{G1}\left(\zeta_2\mathbf{d}_2+\left(1+\zeta_2\right)\mathbf{e}_{2}^\text{DA}\right)
		\end{bmatrix}\in\mathbb{R}^{p_1+p_2},\\
		\mathbf{w}_x^T &= \begin{bmatrix}
			\mathbf{w}^T_1&&\mathbf{w}_2^T
		\end{bmatrix}\in\mathbb{R}^{p_1+p_2}, \label{eq:selectionVector2}\\
		q_x &= e_{\text{s,F}}^\text{q}\frac{1}{G_1G_2}\in\mathbb{R}
	\end{align*}
	in the output signal above results in 
	\begin{equation}
		y_x = \beta x_\text{in} - \mathbf{w}_x^T\boldsymbol{\varphi}_0+q_x,
		\label{eq:outputSignal5}
	\end{equation}
	with a scaling factor $\beta$ and a non-ideal term $\mathbf{w}_x^T\boldsymbol{\varphi}_0$. The weighted digitization error of the final stage is renamed to $q_x$, as the overall digitization is typically referred to as quantization noise. Note that, the derivation above can easily be extended to an arbitrary number of stages. Thus, for the remainder of this work (\ref{eq:outputSignal5}) is not restricted to $n=2$. 
	Analyzing (\ref{eq:outputSignal5}) reveals that the considered non-idealities affect the \ac{adc} output signal in two different ways. First, an overall gain mismatch is introduced with the factor $\beta$, which depends on a combination of all relative gain mismatches. However, scaling does not cause any nonlinear distortion, thus it is not of particular interest in this work. Second, all remaining non-ideal effects are summarized in the vector $\boldsymbol{\varphi}_0$. This suggests to apply $\mathbf{w}_x^T\boldsymbol{\phi}$ as a correction term, with $\boldsymbol{\phi}$ as the calibration parameters resulting in the post-corrected \ac{adc} output
	\begin{equation}
		y_x^\text{c} = \beta x_\text{in}+{\mathbf{w}}_x^T\left({\boldsymbol{\phi}}-{\boldsymbol{\varphi}}_0\right)+q_x.
	\end{equation} 
	\\
	The presented post-correction term may already be used for calibration. Nevertheless, as the relative gain mismatches contribute to each \ac{adc} output value weighted with the corresponding element of $\mathbf{d}_i$, it can be shown that the convergence times of the algorithms in the Sections\,\ref{se:HEC} and \ref{se:XHEC} are significantly improved by incorporating this information into $\mathbf{w}_x$. Therefore, the selection subvectors are rewritten as 
	\begin{equation}
		\left[\bar{\mathbf{w}}_i\right]_j = \begin{cases}\sum_{l=1}^{i}x_{\text{s},l}\prod_{m=1}^{i-l}G_m&\text{if $j = 1$}\\
			\left[\mathbf{w}_i\right]_j&\text{else,}
		\end{cases}
		\label{eq:transselv}
	\end{equation} 
	where $x_{s,l}$ introduces the required information. The subvector $\boldsymbol{\varphi}_{0,i}$ of the parameter vector $\boldsymbol{\varphi}_0$, which corresponds to the $i$th \ac{adc} stage, transforms accordingly as
	\begin{equation*}
		\left[\bar{\boldsymbol{\varphi}}_{0,i}\right]_j = \begin{cases}
			\frac{1}{ \sum_{l=1}^{i}\left[\mathbf{d}_l\right]_1\prod_{m=1}^{i-l}G_m}\left[\boldsymbol{\varphi}_{0,i}\right]_1 &\text{if $j = 1$}\\
			\left[\boldsymbol{\varphi}_{0,i}\right]_j-\frac{\sum_{l=1}^{i}\left[\mathbf{d}_l\right]_j\prod_{m=1}^{i-l}G_m}{\sum_{l=1}^{i}\left[\mathbf{d}_l\right]_1\prod_{m=1}^{i-l}G_m}\left[\boldsymbol{\varphi}_{0,i}\right]_1&\text{else.}
		\end{cases}
	\end{equation*}
	The post-corrected \ac{adc} output signal with transformed non-idealities results in 
	\begin{equation}
		y_x^\text{c} = \beta x_\text{in}+\bar{\mathbf{w}}_x^T\left(\tilde{\boldsymbol{\theta}}-\bar{\boldsymbol{\varphi}}_0\right)+q_x.
		\label{eq:outputSignalCalibrated}
	\end{equation} 
	Choosing transformed calibration parameters $\tilde{\boldsymbol{\theta}}$ eliminates the non-idealities in (\ref{eq:outputSignalCalibrated}), such that only the overall gain mismatch remains. If necessary, the overall gain mismatch can be removed in an additional processing step. As discussed for (\ref{eq:outputSignal3}), the first few stages primarily contribute to the \ac{adc} nonlinearity. Therefore, using only these stages for calibration reduces the complexity without a severe performance degradation. As a consequence, the selection vector and the parameter vector are segmented in the subvectors $\bar{\mathbf{w}}_x^T = \begin{bmatrix}
		\tilde{\mathbf{h}}_x^T &\mathbf{m}_x^T\end{bmatrix}$ and $\bar{\boldsymbol{\varphi}}_0 = \begin{bmatrix}
		\tilde{\boldsymbol{\theta}}_0 &\boldsymbol{\xi}_0\end{bmatrix}$, which correspond to included and excluded stages by the correction model, respectively. The  selection subvector $\tilde{\mathbf{h}}_x^T = \begin{bmatrix}
		\tilde{\mathbf{h}}_1^T&\tilde{\mathbf{h}}_2^T&\ldots&\tilde{\mathbf{h}}_q^T
	\end{bmatrix}$, with $\tilde{\mathbf{h}}_i =\bar{\mathbf{w}}_i$, introduces the first $q$ stages to the post-correction model. The post-corrected output signal follows to
	\begin{equation}
		y_x^\text{c} = y_x+\tilde{\mathbf{h}}_x^T\tilde{\boldsymbol{\theta}} = \beta x_\text{in}+\tilde{\mathbf{h}}_x^T\left(\tilde{\boldsymbol{\theta}}-\tilde{\boldsymbol{\theta}}_0\right)-\mathbf{m}_x^T\boldsymbol{\xi}_0+q_x.
	\end{equation}
	As can be seen in the equation above,  $\tilde{\mathbf{h}}_x^T\tilde{\boldsymbol{\theta}}$ can correct for the first $q$ stages while the less significant stages remain in the post-corrected output signal via the term $\mathbf{m}_x^T\boldsymbol{\xi}_0$. \\
	As calibration requires multiple output samples, a sample index $k$ is introduced for the corresponding variables, e.g., $y_x[k]$ or $\tilde{\mathbf{h}}_x[k]$. To identify a unique parameter vector $\tilde{\boldsymbol{\theta}}$ for different output samples $y_x[k]$, the post-correction matrix $\tilde{\mathbf{H}}_{x}[k]$ in 
	\begin{equation}
		\begin{bmatrix}
			y_x^c[0]\\y_x^c[1]\\\vdots\\y_x^c[k]
		\end{bmatrix} = \begin{bmatrix}
			y_x[0]\\y_x[1]\\\vdots\\y_x[k]
		\end{bmatrix}+\underbrace{\begin{bmatrix}
				\tilde{\mathbf{h}}_x^T[0]\\\tilde{\mathbf{h}}_x^T[1]\\\vdots\\\tilde{\mathbf{h}}_x^T[k]
		\end{bmatrix}}_{\tilde{\mathbf{H}}_{x}[k]}\tilde{\boldsymbol{\theta}}
	\end{equation} 
	has to be of full rank. However, due to the interleaved architecture of pipelined \acp{adc}, one arbitrary parameter of each stage can be represented by an offset in all parameters of the consecutive stage. Thus, the columns of $\tilde{\mathbf{H}}_x[k]$ become linearly dependent. To remove this dependency, one arbitrary entry of each selection subvector $\tilde{\mathbf{h}}_i$, except in the last stage, is eliminated. In this work, always the $p_i$th entry is neglected. Finally, the post-corrected output signal may be written as 
	\begin{equation}
		y_x^c = y_x + \mathbf{h}_x^T\boldsymbol{\theta}, 
		\label{eq:postcorrMod}
	\end{equation}
	with $\mathbf{h}_x^T = \begin{bmatrix}
		\mathbf{h}^T_1&\mathbf{h}^T_{2} &\ldots&\tilde{\mathbf{h}}_q^T
	\end{bmatrix}$ as the reduced selection vector and $\left[\mathbf{h}_i\right]_j = [\tilde{\mathbf{h}}_i]_j$, for $j = 1\ldots p_i-1$. Furthermore, the new parameter vector reads as $\boldsymbol{\theta}^T = \begin{bmatrix}
		{\boldsymbol{\theta}}^T_1 &\boldsymbol{\theta}^T_{2}&\ldots&{\boldsymbol{\theta}}_q^T
	\end{bmatrix}$ with 
	\begin{equation}
		\left[\boldsymbol{\theta}_i\right]_j = \begin{cases}
			[\tilde{\boldsymbol{\theta}}_1]_j&\text{if $i = 1$}\\
			[\tilde{\boldsymbol{\theta}}_i]_j+[\tilde{\boldsymbol{\theta}}_{i-1}]_{p_i}&\text{if $i = 2\ldots q$.}
		\end{cases}
	\end{equation}
	The same parameter reduction may analogously be applied to the system model derived from (\ref{eq:outputSignal5}), such that the true non-idealities $\tilde{\boldsymbol{\theta}}_0$ can be represented by a modified vector $\boldsymbol{\theta}_0$ resulting in the \ac{adc} output 
	\begin{equation}
		y_x = \beta x_\text{in}-\mathbf{h}_x^T\boldsymbol{\theta}_0-\mathbf{m}_x^T\boldsymbol{\xi}_0+q_x.
		\label{eq:adcouptut_fin}
	\end{equation}
	\section{HOMOGENEITY ENFORCED CALIBRATION}
	\label{se:HEC}
	In this section, the fundamentals of the \ac{hec} approach from \cite{1_HEC} are briefly reviewed, followed by a detailed analysis of limitations of the algorithm. 
	\begin{figure}[!t]
		\centering
		\includegraphics{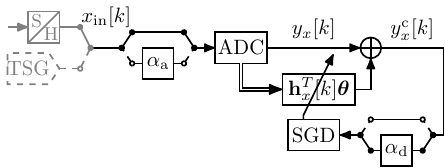}
		\caption{Block diagram of the \ac{hec} approach including two possible signal sources, such as a \ac{tsg} (dashed gray) and an arbitrary input (solid gray), e.g., the signal of the application at hand (modified from \cite{1_HEC}).}
		\label{fig:hec}
	\end{figure}
	\subsection{HEC APPROACH FUNDAMENTALS}
	The \ac{hec} approach aims to enforce the homogeneity condition in (\ref{eq_hom}). Therefore, an input sample $x_{\text{in}}[k]$ is fed into the \ac{adc} twice, whereas for the second time, it is scaled with an analog scaling factor $\alpha_\text{a}$. As can be seen in Fig.~\ref{fig:hec}, possible sources for the input signal are a \ac{tsg} or an arbitrary input, e.g., the signal of the application at hand\footnote{Note that in systems such as, e.g., radar sensors, a \ac{tsg}\cite{fujibayashi_76-_2017} as well as scaling circuits are often already present, as they are used for test or monitoring purposes.}. If an arbitrary input is utilized, an additional \ac{sh} block is required. This \ac{sh} block enables to inject the same input sample twice into the \ac{adc}. The corresponding pair of output samples is denoted by $y_x[k]$ and $y_{\alpha_\text{a} x}[k]$, representing the \ac{adc} output of the non-scaled and scaled input, respectively. Next, $y_x[k]$ is scaled in the digital domain with the scaling factor $\alpha_\text{d}$ and compared with $y_{\alpha_ \text{a}x}[k]$. As a non-ideal \ac{adc} naturally violates the homogeneity condition, these two output values do not match for all possible input values. As introduced in \cite{1_HEC}, the squared difference of these two output signals, in combination with the post-correction model from (\ref{eq:postcorrMod}) yields the cost function
	\begin{equation*}
		J_0[k] = \left(y_{\alpha_\text{a} x}[k]+\mathbf{h}_{\alpha_\text{a} x}^T[k]\boldsymbol{\theta}-\alpha_\text{d}\left(y_x[k]+\mathbf{h}_x^T[k]\boldsymbol{\theta}\right)\right)^2,
		\label{eq:hecCostFun}
	\end{equation*}
	which may be minimized by several standard parameter identification methods \cite{kay_estimation_theory,Kaczmarz,ALS1,Phd_Thesis_OL,diniz1997adaptive,SGD}. In our previous work, we derived the \ac{hec} \ac{sgd} approach, as it supports the low computational complexity of the Kaczmarz algorithm but reduces the total number of samples to be stored. Furthermore, it has been shown that the \ac{hec} \ac{sgd} approach converges in the mean towards the Wiener solution
	\begin{equation}
		\boldsymbol{\theta}_\text{W} = -\mathbf{R}_{\mathbf{hh}}^{-1}\mathbf{r}_{\mathbf{h}y},
		\label{eq:mmseSolution0}
	\end{equation}
	with $\mathbf{R}_{\mathbf{hh}} = \E{\Delta \mathbf{h}[k]\Delta\mathbf{h}^T[k]}{}$, $\mathbf{r}_{\mathbf{h}y} = \E{\Delta\mathbf{h}[k]\Delta y[k]}{}$, $\Delta y[k] = y_{\alpha_\text{a} x}[k]-\alpha_\text{d} y_x[k]$, $\Delta\mathbf{h}[k] = \mathbf{h}_{\alpha_\text{a} x}[k]-\alpha_\text{d} \mathbf{h}_x[k]$, and $\E{\cdot}{}$ denoting the expectation operator.
	\\Although \cite{1_HEC} highlights the benefits of the \ac{hec} approach by means of high calibration performance without a precisely known test signal, it is also subject to certain limitations. In the following, these limitations are investigated in detail.  
	\subsection{HEC APPROACH LIMITATIONS}
	\label{se:limitations}
	For simplicity, the limitations are only analyzed based on the \ac{hec} Wiener solution. However, as the \ac{hec} \ac{sgd} approach converges in the mean towards the \ac{hec} Wiener solution, all presented findings are also valid for the \ac{hec} \ac{sgd} approach.\\
	Considering the input sample
	\begin{equation}
		x_\text{in}[k] = x_\text{d}[k]+n_x[k]
		\label{eq:xinAndnoise}
	\end{equation}
	consisting of a deterministic signal $x_\text{d}[k]$ and noise $n_x[k]$, the two \ac{adc} outputs with the original and the scaled input, respectively, read as 
	\begin{equation}
		\begin{split}
			y_x[k] &= \beta \left(x_\text{d}[k]+n_x[k]\right)-\mathbf{h}_x^T[k]\boldsymbol{\theta}_0-\mathbf{m}_x^T[k]\boldsymbol{\xi}_0+q_x[k] \label{eq:xoutADCandNoise1}
		\end{split}
	\end{equation}
	and
	\begin{equation}
		\begin{split}
			y_{\alpha_\text{a}x}[k] &= \beta \left(\alpha_{\text{a}}x_\text{d}[k]+n_{\alpha_{\text{a}}x}[k]\right)-\mathbf{h}_{\alpha_\text{a}x}^T[k]\boldsymbol{\theta}_0-  \mathbf{m}_{\alpha_\text{a}x}^T[k]\boldsymbol{\xi}_0+q_{\alpha_\text{a}x}[k].
		\end{split} \label{eq:xoutADCandNoise2}
	\end{equation}
	With the two output signals from above and by introducing a scaling factor mismatch as $\alpha_\text{d} = \alpha_\text{a}-\delta$, 
	\begin{equation}
		\Delta y[k] = \beta\Delta n[k]-\Delta \mathbf{h}^T[k]\boldsymbol{\theta}_0-\Delta\mathbf{m}^T[k]\boldsymbol{\xi}_0+\Delta q[k], 
		\label{eq:dy}
	\end{equation}
	where $\Delta n[k] =n_{\alpha_\text{a} x}[k]-\left(\alpha_\text{a}-\delta\right)n_x[k]$, $\Delta \mathbf{m}[k] = \mathbf{m}_{\alpha_\text{a} x}[k]-\left(\alpha_\text{a}-\delta\right)\mathbf{m}_x[k]$, and  $\Delta q[k] =q_{\alpha_\text{a} x}[k]-\left(\alpha_\text{a}-\delta\right)q_x[k]$.
	Substituting \eqref{eq:dy} into \eqref{eq:mmseSolution0}, yields
	\begin{equation}
		\begin{split}
			{{\boldsymbol{\theta}}}_\text{W} =&\text{\space} \boldsymbol{\theta}_0+\mathbf{R}_{\mathbf{hh}}^{-1}\left(\mathbf{R}_{\mathbf{hm}}\boldsymbol{\xi}_0-\delta\mathbf{r}_{\mathbf{h}x}-\mathbf{r}_{\mathbf{h}q}-\mathbf{r}_{\mathbf{h}n}\right),
		\end{split}
		\label{eq:mmseSolution1}
	\end{equation}
	with $\mathbf{R}_\mathbf{hm}=\E{\Delta\mathbf{h}[k]\Delta\mathbf{m}^T[k]}{}$,  $\mathbf{r}_{\mathbf{h}x} = \E{\Delta\mathbf{h}[k]\beta x_\text{d}[k]}{}$, $\mathbf{r}_{\mathbf{h}q} = \E{\Delta\mathbf{h}[k]\Delta q[k]}{}$, and $\mathbf{r}_{\mathbf{h}n} = \E{\Delta\mathbf{h}[k]\beta\Delta n[k]}{}$.
	Finally, assuming that $\Delta \mathbf{h}[k]$ and $\Delta q[k]$ are uncorrelated results in 
	\begin{equation}
		{{\boldsymbol{\theta}}}_\text{W} = \boldsymbol{\theta}_0+\mathbf{R}_{\mathbf{hh}}^{-1}\mathbf{R}_\mathbf{hm}\boldsymbol{\xi}_0-\mathbf{R}_{\mathbf{hh}}^{-1}\mathbf{r}_{\mathbf{h}n}-\delta\mathbf{R}_{\mathbf{hh}}^{-1}\mathbf{r}_{\mathbf{h}x}.
		\label{eq:mmseSolution2}
	\end{equation}
	As can be seen, the \ac{hec} Wiener solution deviates from the true \ac{adc} non-idealities due to three effects, which are excluded non-idealities, analog input noise, and scaling factor mismatch.  
	\paragraph*{Exluded Non-Idealities}
	The first bias term is caused by non-idealities which are excluded from the post-correction, such that $\boldsymbol{\xi}_0 \neq \mathbf{0}$. In \eqref{eq:mmseSolution2}, this bias term appears because of a non-zero cross-correlation matrix $\mathbf{R}_\mathbf{hm}$. This cross-correlation matrix is mainly non-zero due to the interleaved architecture of pipelined \acp{adc}, where one parameter of each stage can be represented by an offset in all parameters of the consecutive stage, as already discussed in Section\,\ref{sec:postcorrmodel}.  
	\paragraph*{Analog Input Noise}
	The second bias term in \eqref{eq:mmseSolution2} is caused by analog input noise according to  $\mathbf{R}_{\mathbf{hh}}^{-1}\mathbf{r}_{\mathbf{h}n}$. Analyzing this expression shows that the cross-correlation vector
	\begin{equation*}
		\mathbf{r}_{\mathbf{h}n} = \beta\E{\left(\mathbf{h}_{\alpha_\text{a} x}[k]-\alpha_\text{d}\mathbf{h}_x[k]\right)\left(n_{\alpha_\text{a} x}[k]-\alpha_\text{d}n_x[k]\right)}{}
	\end{equation*} 
	may be written as 
	\begin{equation}
		\mathbf{r}_{\mathbf{h}n} = \beta\left(\E{\mathbf{h}_{\alpha_\text{a} x}[k]n_{\alpha_\text{a} x}[k]}{}+\alpha_\text{d}^2\E{\mathbf{h}_x[k]n_x[k]}{}\right)
		\label{eq:crosscorbias}
	\end{equation}
	since $\mathbf{h}_{\alpha_\text{a} x}$ and $n_x[k]$ as well as $\mathbf{h}_x[k]$ and $n_{\alpha_\text{a} x}[k]$ are uncorrelated. As derived in Section\,\ref{sec:postcorrmodel}, the selection vectors $\mathbf{h}_x[k]$ and $\mathbf{h}_{\alpha_\text{a} x}[k]$ depend on the current input samples and therefore the current noise samples $n_x[k]$ and $n_{\alpha_\text{a} x}[k]$, respectively. Hence, noise samples induce changes in the selection vectors, which contribute to the cross-correlation in \eqref{eq:crosscorbias}. Consequently, the impact of analog input noise depends on the noise power and on the number of stages which are considered for calibration. 
	\paragraph*{Scaling Factor Mismatch}
	The third bias term is caused by mismatching analog and digital scaling factors $\alpha_\text{a}$ and $\alpha_\text{d}$ resulting in $\delta \neq 0$. As can be seen in \eqref{eq:mmseSolution2}, this introduces the term $\delta\mathbf{R}_{\mathbf{hh}}^{-1}\mathbf{r}_{\mathbf{h}x}$ weighted with the scaling factor mismatch $\delta$. This term depends on the strong cross-correlation between $\Delta\mathbf{h}[k]$ and the input signal $x_\text{d}[k]$. Based on \ac{pvt} variations in a real-world on-chip implementation, this may be considered as the main limitation of the \ac{hec} approach which reduces its calibration abilities. To cope with this performance decrease, an enhanced version of the \ac{hec} approach is introduced in the following section. 
	\section{BI-LINEAR HOMOGENEITY ENFORCED CALIBRATION}
	\label{se:XHEC}
	As shown in \cite{1_HEC}, the \ac{hec} approach performs well in an ideal environment. However, the last Section~pointed out that its performance is highly sensitive to a mismatch between the analog and the digital scaling factors. To deal with this scaling factor mismatch, the fundamental inequality of the \ac{hec} approach is modified by introducing a new parameter, which can compensate for the scaling factor mismatch. Specifically, the extended inequality is written as 
	\begin{equation*}
		y_{\alpha_\text{a}x}[k]+\mathbf{h}_{\alpha_\text{a}x}^T[k]\boldsymbol{\theta}_\text{NL}\neq\left(\alpha_\text{d}+\theta_\alpha\right)\left(y_x[k]+\mathbf{h}_x^T[k]\boldsymbol{\theta}_\text{NL}\right),
		\label{eq:hecInequality2}
	\end{equation*}
	with $\theta_\alpha$ as a new parameter and the renamed parameter vector $\boldsymbol{\theta}$ as $\boldsymbol{\theta}_\text{NL}$. Computing the difference of both sides yields 
	\begin{equation}
		\begin{split}
			e[k] =&\text{\space} y_{\alpha_\text{a}x}[k]+\mathbf{h}_{\alpha_\text{a}x}^T[k]\boldsymbol{\theta}_\text{NL}-\left(\alpha_\text{d}+\theta_\alpha\right)\left(y_x[k]+\mathbf{h}_x^T[k]\boldsymbol{\theta}_\text{NL}\right).
		\end{split}
		\label{eq:homogeneityError}
	\end{equation}
	A closer look shows that the error is no longer linear in all parameters. Caused by the newly introduced parameter $\theta_\alpha$, the error consists of a linear part in $\boldsymbol{\theta}_\text{NL}$ and $\theta_\alpha$, as well as of a part that is bi-linear in $\boldsymbol{\theta}_\text{NL}$ and $\theta_\alpha$.\\
	In general, bi-linear models are well known in literature since they can cope with a large class of nonlinear systems while they behave, to some extent, similar as linear systems \cite{3_EST_AnOverviewOfBilinearSystemTheoryAndApplications,4_EST_BilinearSystemsAnAppealingClassOf}. In \cite{3_EST_AnOverviewOfBilinearSystemTheoryAndApplications,4_EST_BilinearSystemsAnAppealingClassOf,5_EST_TheLeastSquaresBasedIterativeAlgorithms,9_EST_PracticallyEfficientNonlinearAcousticEchoCanceller,10_EST_NonlinearActiveNoiseControl,11_EST_LowComplexityNonlinearAdaptivFilter,1_EST_AnalysisOfAnLMSAlgorithmForBilinearForms,2_EST_OnTheIdentificationOfBilinearFormsWithTheWienerFilter,6_EST_AdaptiveFilteringForTheIdent,7_EST_IdentificationOfBilinearFormsWithTheKalman,8_EST_IdentOfLinearAndBilinearSystems}, different methods for parameter estimation of bi-linear models are introduced, e.g., the method of least squares, the Wiener filter, a recursive least squares estimator, and a least mean squares filter. They all consider a system model, that is bi-linear in the input \cite{3_EST_AnOverviewOfBilinearSystemTheoryAndApplications,4_EST_BilinearSystemsAnAppealingClassOf,5_EST_TheLeastSquaresBasedIterativeAlgorithms,9_EST_PracticallyEfficientNonlinearAcousticEchoCanceller,10_EST_NonlinearActiveNoiseControl,11_EST_LowComplexityNonlinearAdaptivFilter}, or is purely bi-linear in the parameters \cite{1_EST_AnalysisOfAnLMSAlgorithmForBilinearForms,2_EST_OnTheIdentificationOfBilinearFormsWithTheWienerFilter,6_EST_AdaptiveFilteringForTheIdent,7_EST_IdentificationOfBilinearFormsWithTheKalman,8_EST_IdentOfLinearAndBilinearSystems}. In contrast to the latter, the error in (\ref{eq:homogeneityError}) is linear and bi-linear in the parameters. Hence, the results derived in \cite{1_EST_AnalysisOfAnLMSAlgorithmForBilinearForms,2_EST_OnTheIdentificationOfBilinearFormsWithTheWienerFilter,6_EST_AdaptiveFilteringForTheIdent,7_EST_IdentificationOfBilinearFormsWithTheKalman,8_EST_IdentOfLinearAndBilinearSystems} cannot be applied in this work. In the following, two novel algorithms that can deal with the mixed linear/bi-linear error in (\ref{eq:homogeneityError}), are derived. 
	\subsection{BL-HEC WIENER SOLUTION}
	Based on the homogeneity error in (\ref{eq:homogeneityError}), the \ac{mse} cost function is given by
	\begin{equation}
		J^\text{MSE}_\text{ext}(\theta_\alpha,\boldsymbol{\theta}_\text{NL}) =\E{e^2[k]}{}.
		\label{eq:mmseCostFunction2}
	\end{equation}
	Note that (\ref{eq:mmseCostFunction2}) is a function of two different parameters, namely $\theta_\alpha$ and $\boldsymbol{\theta}_\text{NL}$. As suggested in \cite{2_EST_OnTheIdentificationOfBilinearFormsWithTheWienerFilter}, the Wiener solution is derived for both parameters separately. First, (\ref{eq:mmseCostFunction2}) is minimized with respect to $\theta_\alpha$ yielding 
	\begin{equation}
		\theta_{\alpha,\text{W}} = \frac{{r}_{yy_\alpha}(\boldsymbol{\theta}_\text{NL})}{{r}_{yy}(\boldsymbol{\theta}_\text{NL})}-\alpha_\text{d}, 
		\label{eq:mmseSolutionAlpha1}
	\end{equation}
	with \begin{align}
		{r}_{yy}(\boldsymbol{\theta}_\text{NL}) &= \E{\left(y_x[k]+\mathbf{h}_x^T[k]\boldsymbol{\theta}_\text{NL}\right)^2}{},\\
		\begin{split}
			{r}_{yy_\alpha}(\boldsymbol{\theta}_\text{NL})  &=  E\left[\left(y_{\alpha_\text{a}x}[k]+\mathbf{h}_{\alpha_\text{a}x}^T[k]\boldsymbol{\theta}_\text{NL}\right)\left(y_x[k]+\mathbf{h}_x^T[k]\boldsymbol{\theta}_\text{NL}\right)\right].
		\end{split}
	\end{align} It should be noted that, the Wiener solution $\theta_{\alpha,\text{W}}$ is a function of the parameter vector $\boldsymbol{\theta}_\text{NL}$ by means that each realization of $\boldsymbol{\theta}_\text{NL}$ results in a different $\theta_{\alpha,\text{W}}$. Second, minimizing (\ref{eq:mmseCostFunction2}) with respect to $\boldsymbol{\theta}_\text{NL}$ leads to 
	\begin{equation}
		\boldsymbol{\theta}_{\text{NL},\text{W}} =-\mathbf{R}_{\mathbf{hh}}^{-1}(\theta_\alpha)\mathbf{r}_{\mathbf{h}y}(\theta_{\alpha}),
		\label{eq:mmseSolutionTheta1}
	\end{equation}
	with 
	\begin{equation}
		\mathbf{R}_{\mathbf{hh}}(\theta_\alpha) = \E{\Delta\mathbf{h}[k,\theta_\alpha]\Delta\mathbf{h}^T[k,\theta_\alpha]}{}
	\end{equation}
	and
	\begin{equation}
		\begin{split}
			\mathbf{r}_{\mathbf{h}y}(\theta_{\alpha}) =&\text{\space} \E{\Delta\mathbf{h}[k,\theta_\alpha]\left( y_{\alpha_\text{a}x}[k]-\left(\alpha_\text{d}+\theta_\alpha\right)y_x[k]\right)}{},
		\end{split}
	\end{equation}
	where $\Delta\mathbf{h}[k,\theta_\alpha] = \mathbf{h}_{\alpha_\text{a}x}[k]-\left(\alpha_\text{d}+\theta_\alpha\right)\mathbf{h}_x[k]$. 
	In this case, the optimum solution is a function of the parameter $\theta_\alpha$. In general, a closed form solution of (\ref{eq:mmseSolutionAlpha1}) and (\ref{eq:mmseSolutionTheta1}) does not exist. As in \cite{2_EST_OnTheIdentificationOfBilinearFormsWithTheWienerFilter}, the optimum solution has to be evaluated by an iterative approach, yielding the \ac{xhec} Wiener solution.\\ To obtain the \ac{xhec} Wiener solution, the parameter vector $\boldsymbol{\theta}_{\text{NL},\text{W}}[0]$ is initialized as the zero vector and inserted into 
	\begin{equation}
		\theta_{\alpha,\text{W}}[m] =\frac{{r}_{yy_\alpha}(\boldsymbol{\theta}_\text{NL,W}[m-1])}{{r}_{yy}(\boldsymbol{\theta}_\text{NL,W}[m-1])}-\alpha_\text{d},
		\label{eq:mmseSolutionAlpha2}
	\end{equation}
	with the iteration index $m = 1\ldots M$. In a second step, the solution of (\ref{eq:mmseSolutionAlpha2}) is used to compute the updated parameter vector 
	\begin{equation}
		\boldsymbol{\theta}_{\text{NL},\text{W}}[m] =-\mathbf{R}_{\mathbf{hh}}^{-1}(\theta_{\alpha,\text{W}}[m])\mathbf{r}_{\mathbf{h}y}(\theta_{\alpha,\text{W}}[m]).
		\label{eq:mmseSolutionTheta2}
	\end{equation}
	Iteratively evaluating (\ref{eq:mmseSolutionAlpha2}) and (\ref{eq:mmseSolutionTheta2}) results in the iterative \ac{xhec} Wiener solution.
	\begin{figure}[!t]
		\centering
		\includegraphics{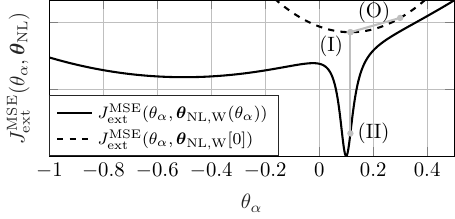}
		\caption{Non-convex \ac{mse} cost function of an exemplary \ac{adc}.}
		\label{fig:nonConvexCost}
	\end{figure} 
	Subsequently, it will be shown that this iterative procedure converges towards a minimum of (\ref{eq:mmseCostFunction2}). \\At first, the \ac{mse} cost function itself in (\ref{eq:mmseCostFunction2}) is analyzed. Therefore, the cost function for one exemplary non-ideal \ac{adc} is illustrated by the solid black line in Fig.~\ref{fig:nonConvexCost}. To obtain this black line, $\boldsymbol{\theta}_\text{NL}$ in (\ref{eq:mmseCostFunction2}) is replaced by the Wiener solution from (\ref{eq:mmseSolutionTheta1}). This yields the \ac{mse} $J^\text{MSE}_\text{ext}(\theta_\alpha,\boldsymbol{\theta}_\text{NL,W}(\theta_\alpha))$, which only depends on $\theta_\alpha$.
	As can clearly be seen in Fig.~\ref{fig:nonConvexCost}, this function is not convex and provides two distinct minima. The global minimum is located at $\theta_\alpha \approx \delta = 0.1$ which corresponds to the simulated scaling factor mismatch between $\alpha_\text{a}$ and $\alpha_\text{d}$ of the example \ac{adc}. A local minimum appears at $\theta_\alpha \approx -\alpha_\text{d} = -0.5$. A closer look on the homogeneity error from \eqref{eq:homogeneityError} in (\ref{eq:mmseCostFunction2}) shows that for $\theta_\alpha = -\alpha_\text{d}$, the second term vanishes, resulting in 
	\begin{equation}
		J_\text{ext}^\text{MSE}(-\alpha_\text{d},\boldsymbol{\theta}_\text{NL}) = \E{\left(y_{\alpha_\text{a}x}[k]+\mathbf{h}_{\alpha_\text{a}x}^T[k]\boldsymbol{\theta}_\text{NL}\right)^2}{}.
	\end{equation}
	Obviously, this local minimum is of no practical interest since it simply minimizes the output of the \ac{adc}. \\
	To show convergence of the iterative \ac{xhec} Wiener solution towards one of these two minima, assume a fixed $\boldsymbol{\theta}_{\text{NL,W}}[0]$. Consequently, the homogeneity error (\ref{eq:homogeneityError}) becomes linear in $\theta_\alpha$. Thus, the temporary cost function $J_\text{ext}^\text{MSE}(\theta_\alpha, \boldsymbol{\theta}_\text{NL,W}[0])$ is convex with a global temporary minimum. This temporary cost function is illustrated by the dashed black line in Fig.~\ref{fig:nonConvexCost}. With $\theta_{\alpha\text{,W}}[1]$ from (\ref{eq:mmseSolutionAlpha2}), the temporary cost function is minimized such that 
	\begin{equation}
		J_\text{ext}^\text{MSE}(\theta_{\alpha,\text{W}}[0], \boldsymbol{\theta}_\text{NL,W}[0])\geq J_\text{ext}^\text{MSE}(\theta_{\alpha,\text{W}}[1], \boldsymbol{\theta}_\text{NL,W}[0]).
		\label{eq:conv1}
	\end{equation}
	This iteration step is illustrated by the points (\Romannum{0})-(\Romannum{1}) in Fig.~\ref{fig:nonConvexCost}. Next, $\theta_{\alpha,\text{W}}[1]$ is fixed resulting in another convex temporary cost function $J_\text{ext}^\text{MSE}(\theta_{\alpha,\text{W}}[1], \boldsymbol{\theta}_\text{NL})$, which is minimized by $\boldsymbol{\theta}_\text{NL,W}[1]$ from (\ref{eq:mmseSolutionTheta2}), thus
	\begin{equation}
		J_\text{ext}^\text{MSE}(\theta_{\alpha,\text{W}}[1], \boldsymbol{\theta}_\text{NL,W}[0])\geq J_\text{ext}^\text{MSE}(\theta_{\alpha,\text{W}}[1], \boldsymbol{\theta}_\text{NL,W}[1]),
		\label{eq:conv2}
	\end{equation}
	as marked by the points  (\Romannum{1})-(\Romannum{2}) in Fig.~\ref{fig:nonConvexCost}.
	Combining (\ref{eq:conv1}) and (\ref{eq:conv2}), one can see that each iteration of the iterative \ac{xhec} Wiener solution has to reduce the \ac{mse} in (\ref{eq:mmseCostFunction2}). Finally, equality in either (\ref{eq:conv1}) or (\ref{eq:conv2}) directly implies convergence of the iterative \ac{xhec} Wiener solution. Fig.~\ref{fig:nonConvexCost} shows the temporary cost function  $J_\text{ext}^\text{MSE}(\theta_\alpha, \boldsymbol{\theta}_\text{NL,W}[0])$ by the dashed black line.\\
	Nevertheless, convergence towards the global minimum strongly depends on the initialization. In this work, $\boldsymbol{\theta}_\text{NL,W}[0]$ is initialized as the zero vector, which shows consistently good results in both simulation and measurements. To provide analytical insight, consider the first step of the iterative \ac{xhec} Wiener solution from (\ref{eq:mmseSolutionAlpha2}). This step yields 
	\begin{equation}
		\theta_{\alpha,\text{W}}[1] = \frac{\E{y_{\alpha_\text{a}x}[k]y_x[k]}{}}{\E{\left(y_x[k]\right)^2}{}}-\alpha_\text{d}, 
	\end{equation}
	with $y_{\alpha_\text{a} x}[k]$ and $y_x[k]$ as the uncalibrated \ac{adc} outputs. Assuming practically relevant \ac{adc} non-idealities results in a \ac{sndr} higher than $30\,\text{dB}$ (see Table\,\ref{tab:comp}). Consequently, the term $\E{\left(y_x[k]\right)^2}{}$ is dominated by the test signal power while noise and distortion from the \ac{adc} non-idealities may be neglected. Although the test signal power in the term $\E{y_{\alpha_\text{a}x}[k]y_x[k]}{}$ is reduced by $\alpha_{\text{a}}$ it is still significantly higher than the contained noise and distortion. Hence, already the first iteration directs the \ac{xhec} Wiener towards the global minimum, which supports the observed convergence behavior in practical scenarios (see Section\,\ref{se:Results}).  \\
	Finally, the global minimum itself is analyzed. With the aid of some minor approximations, the minimum can be written in a closed form. Analogously to the derivation of \eqref{eq:mmseSolution2} in Section\,\ref{se:limitations}, inserting \eqref{eq:xoutADCandNoise1} and \eqref{eq:xoutADCandNoise2} into  \eqref{eq:mmseSolutionTheta1}, considering $\alpha_\text{d} =\alpha_\text{a}-\delta$, and with $\Delta\mathbf{h}[k,\theta_\alpha]$ and $\Delta q[k,\theta_\alpha]$ being uncorrelated, allows rewriting the \ac{xhec} Wiener solution as
	\begin{equation}
		\begin{split}
			\boldsymbol{\theta}_\text{NL,W} =& \text{\space} \boldsymbol{\theta}_0+\mathbf{R}_{\mathbf{hh}}^{-1}(\theta_\alpha)\left(\mathbf{R}_\mathbf{hm}(\theta_\alpha)\boldsymbol{\xi}_0-\mathbf{r}_{\mathbf{h}n}(\theta_\alpha)-\left(\delta-\theta_{\alpha}\right)\mathbf{r}_{\mathbf{h}x}(\theta_\alpha)\right),
		\end{split}
		\label{eq:blwiener}
	\end{equation}
	where $\mathbf{R}_\mathbf{hm}(\theta_\alpha)=\E{\Delta\mathbf{h}[k,\theta_\alpha]\Delta\mathbf{m}[k,\theta_\alpha]}{}$, $\Delta\mathbf{m}[k,\theta_\alpha] = \mathbf{m}_{\alpha_\text{a} x}[k]-\left(\alpha_\text{a}-\delta+\theta_\alpha\right)\mathbf{m}_x[k]$, $\mathbf{r}_{\mathbf{h}x}(\theta_\alpha) = \E{\Delta\mathbf{h}[k,\theta_\alpha]\beta x_\text{d}[k]}{}$, $\mathbf{r}_{\mathbf{h}n}(\theta_\alpha) = \E{\Delta\mathbf{h}[k,\theta_\alpha]\beta\Delta n[k,\theta_\alpha]}{}$, $\Delta n[k,\theta_\alpha] =n_{\alpha_\text{a} x}[k]-\left(\alpha_\text{a}-\delta+\theta_{\alpha}\right)n_x[k]$, and $\Delta q[k,\theta_\alpha] = q_{\alpha_{\text{a}}x}[k]-\left(\alpha_{\text{a}}-\delta+\theta_\alpha\right)q_x[k]$. As with the \ac{hec} approach in \eqref{eq:mmseSolution2}, \eqref{eq:blwiener} shows three bias terms caused by excluded non-idealities, analog input noise and the scaling factor mismatch. By considering a sufficiently high \ac{snr}, and including all significant stages to the post-correction model, \eqref{eq:blwiener} reduces to 
	\begin{equation}
		\boldsymbol{\theta}_\text{NL,W} \approx  \boldsymbol{\theta}_0+\left(\delta-\theta_{\alpha}\right)\mathbf{R}_{\mathbf{hh}}^{-1}(\theta_\alpha)\mathbf{r}_{\mathbf{h}x}(\theta_\alpha).
		\label{eq:blwienersimp}
	\end{equation}
	Next \eqref{eq:xoutADCandNoise1}, \eqref{eq:xoutADCandNoise2}, and $\alpha_{\text{d}}= \alpha_{\text{a}}-\delta$ are inserted into the homogeneity error in \eqref{eq:homogeneityError}, thus
	\begin{equation}
		\begin{split}
			e[k] =&\text{\space} \beta\left(\delta-\theta_\alpha\right)x_\text{d}[k]+\beta\Delta n[k,\theta_{\alpha}]-\Delta\mathbf{h}^T[k,\theta_\alpha]\boldsymbol{\theta}_0-\Delta\mathbf{m}^T[k,\theta_{\alpha}]\boldsymbol{\xi}_0+\Delta q[k,\theta_{\alpha}]+\Delta\mathbf{h}^T[k,\theta_\alpha]\boldsymbol{\theta}_\text{NL}.
		\end{split}
	\end{equation}
	Inserting \eqref{eq:blwienersimp} in this error and with the same assumption from above, it reduces to  
	\begin{equation}
		\begin{split}
			e[k] \approx&\text{\space} \left(\delta-\theta_{\alpha}\right)\left(\beta x_\text{d}[k]-\vphantom{\left(\Delta\mathbf{h}(\theta_\alpha)\right)^T}\Delta\mathbf{h}^T[k,\theta_\alpha]\mathbf{R}_{\mathbf{hh}}^{-1}(\theta_\alpha)\mathbf{r}_{\mathbf{h}x}(\theta_{\alpha})\right)+\Delta q[k,\theta_\alpha]
		\end{split}
	\end{equation}
	Utilizing this error to compute the \ac{mse} cost function shown in Fig.~\ref{fig:nonConvexCost} results in 
	\begin{equation}	
		\begin{split}
			J^\text{MSE}_\text{ext}(\theta_\alpha,\boldsymbol{\theta}_\text{NL,W}(\theta_\alpha)) \approx&\text{\space} \sigma_{q}^2+\left(\delta-\theta_{\alpha}\right)^2\left(\sigma_{x_\text{d}}^2-\vphantom{\left(\mathbf{r}_{\mathbf{h}x}(\theta_\alpha)\right)^T}\hspace{-2em}\mathbf{r}_{\mathbf{h}x}^T(\theta_\alpha)\mathbf{R}_{\mathbf{hh}}^{-1}(\theta_\alpha)\mathbf{r}_{\mathbf{h}x}(\theta_\alpha)\right),
		\end{split}
		\label{eq:minMSE}
	\end{equation}
	with $\sigma_q^2 = \E{\left(\Delta q[k,\theta_\alpha]\right)^2}{}$ and $\sigma_{x_\text{d}}^2 = \E{\left(\beta x_\text{d}[k]\right)^2}{}$.
	Considering a high-resolution \ac{adc}, the quantization noise $\sigma_q^2$ may be neglected, thus the global minimum from \eqref{eq:minMSE} is clearly given by $\theta_\alpha = \delta$. Reinserting this result in \eqref{eq:blwienersimp} shows that the \ac{xhec} Wiener solution yields a good approximation of the true scaling factor mismatch in addition to the true \ac{adc} non-idealities. Nevertheless, it requires knowledge of the statistics which is unknown in practice and therefore has to be estimated additionally. Furthermore, in (\ref{eq:mmseSolutionTheta2}) the matrix $\mathbf{R}_{\mathbf{ hh}}(\theta_\alpha[m])$ needs to be inverted in each iteration step, which adds to the computational complexity of the \ac{xhec} Wiener solution. Thus this solution is not intended for on-chip implementation but serves as a performance reference for the rest of this work. To achieve the high calibration performance of the \ac{xhec} Wiener filter with a significantly lower computational effort, an adaptive approach is presented in the following section.
	\subsection{BL-HEC SGD APPROACH}
	To derive the \ac{xhec} \ac{sgd} approach, the homogeneity error (\ref{eq:homogeneityError}) is used in the instantaneous squared error cost function as $J_\text{ext}^\text{SGD}[k] = e^2[k]$. The gradient of $J_\text{ext}^\text{SGD}[k]$ with respect to $\theta_\alpha$ yields the parameter update
	\begin{equation}
		\begin{split}
			\theta_\alpha[k] = &\text{\space}\theta_\alpha[k-1]+\mu_\alpha\left(y_x[k]+\mathbf
			{h}_x^T[k]\boldsymbol{\theta}_\text{NL}[k-1]\right)e_\alpha[k]
		\end{split}
		\label{eq:alphaUpdate}
	\end{equation}
	with the apriori error 
	\begin{equation}
		\begin{split}
			e_\alpha[k] =&\text{\space} y_{\alpha_\text{a} x}[k]+\mathbf{h}_{\alpha_\text{a} x}^T[k]\boldsymbol{\theta}_\text{NL}[k-1]-\left(\alpha_\text{d}+\theta_\alpha[k-1]\right)\left(y_x[k]+\mathbf
			{h}_x^T[k]\boldsymbol{\theta}_\text{NL}[k-1]\right)
		\end{split}
		\label{eq:apriorierr_alpha}
	\end{equation}
	and the step-size $\mu_\alpha$.
	Analogously, the gradient of $J_\text{ext}^\text{SGD}[k]$ with respect to $\boldsymbol{\theta}_\text{NL}$ yields the parameter update 
	\begin{equation}
		\begin{split}
			\boldsymbol{\theta}_\text{NL}[k] =&\text{\space} \boldsymbol{\theta}_\text{NL}[k-1]-\mu_\text{NL}\left(\mathbf{h}_{\alpha_\text{a}x}[k]-\left(\alpha_\text{d}+\theta_\alpha[k]\right)\mathbf{h}_x[k]\right)e_\text{NL}[k]
		\end{split}
		\label{eq:nlUpdate}
	\end{equation}
	with the apriori error 
	\begin{equation}
		\begin{split}
			e_\text{NL}[k] =&\text{\space} y_{\alpha_\text{a} x}[k]+\mathbf{h}_{\alpha_\text{a} x}^T[k]\boldsymbol{\theta}_\text{NL}[k-1]-\left(\alpha_\text{d}+\theta_\alpha[k]\right)\left(y_x[k]+\mathbf
			{h}_x^T[k]\boldsymbol{\theta}_\text{NL}[k-1]\right)
		\end{split}
		\label{eq:apriorierr_nl}
	\end{equation}
	and the step-size $\mu_\text{NL}$. 
	Similarly to the \ac{xhec} Wiener solution, both update equations depend on each other, by means they have to be evaluated alternately. 
	\\Note that in \cite{1_HEC}, it was shown that the \ac{hec} approach requires $\sum_{i=1}^q p_i - \left(q-1\right)$ multiplications in each update step. The computational complexity of the \ac{xhec} \ac{sgd} approach differs from the original \ac{hec} approach by the scalar valued update equation (\ref{eq:alphaUpdate}). As the calibrated output signals are also needed for the original \ac{hec} approach, computing (\ref{eq:alphaUpdate}) only adds 3 scalar multiplications per update step.
	\subsection{STABILITY}
	In the following, stability boundaries of the \ac{xhec} \ac{sgd} step-sizes $\mu_\alpha$ and $\mu_\text{NL}$ will be derived.\\
	To show stability, the two \ac{xhec} \ac{sgd} update equations from \eqref{eq:alphaUpdate} and \eqref{eq:nlUpdate} are analyzed separately. Therefore, an arbitrary but constant $\boldsymbol{\theta}_\text{NL}[k] = \boldsymbol{\theta}_\text{NL}$ is used to compute the \ac{xhec} Wiener solution $\theta_{\alpha,\text{W}}(\boldsymbol{\theta}_\text{NL})$ from \eqref{eq:mmseSolutionAlpha1}. This constant \ac{xhec} Wiener solution enables to introduce an error variable $v_\alpha[k] = \theta_\alpha[k]-\theta_{\alpha,\text{W}}(\boldsymbol{\theta}_\text{NL})$. Subtracting $\theta_{\alpha,\text{W}}(\boldsymbol{\theta}_\text{NL})$ from both sides of (\ref{eq:alphaUpdate}), computing the expected value of the result, assuming that $\theta_\alpha[k-1]$ is statistically independent of $y_x[k]+\mathbf{h}_x^T[k]\boldsymbol{\theta}_\text{NL}[k-1]$, and still considering an arbitrary but constant $\boldsymbol{\theta}_\text{NL}[k] = \boldsymbol{\theta}_\text{NL}$ results in 
	\begin{equation}
		\begin{split}
			\E{v_\alpha[k]}{} =&\text{\space} \E{v_\alpha[k-1]}{}+\mu_\alpha\left({r}_{yy_\alpha}(\boldsymbol{\theta}_\text{NL})-\E{\left(\alpha_\text{d}+\theta_\alpha[k-1]\right)}{}{r}_{yy}(\boldsymbol{\theta}_\text{NL})\right).
		\end{split}
		\label{eq:valpha}
	\end{equation}
	Separating $\alpha_\text{d}$ from (\ref{eq:mmseSolutionAlpha1}) and inserting this result into (\ref{eq:valpha}) yields 
	\begin{equation}
		\E{v_\alpha[k]}{} = \left(1-\mu_\alpha{r}_{yy}(\boldsymbol{\theta}_\text{NL})\right)\E{v_\alpha[k-1]}{}.
		\label{eq:convAlpha}
	\end{equation}
	\begin{figure}[!t]
		\centering
		\includegraphics{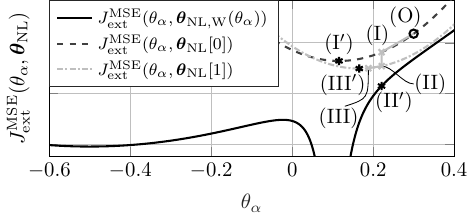}
		\caption{Exemplary \ac{xhec} \ac{sgd} convergence in the \ac{mse} cost function with the black and the gray markers representing the \ac{xhec} Wiener and the \ac{xhec} \ac{sgd} updates, respectively.}
		\label{fig:nonConvexCost_lms}
	\end{figure} 
	Clearly, with the step-size chosen such that $0 < \mu_\alpha < 2/{r}_{yy}(\boldsymbol{\theta}_\text{NL})$, the mean of the error $\E{v_\alpha[k]}{}$ converges towards zero, and consequently $\theta_\alpha[k]$ converges in the mean towards the constant \ac{xhec} Wiener solution. To interpret this behavior with respect to the \ac{mse} cost function, the constant $\boldsymbol{\theta}_\text{NL}$ as well as $\E{\theta_\alpha[k]}{}$ are inserted into \eqref{eq:mmseCostFunction2}, resulting in 
	\begin{equation}
		\begin{split}
			J^\text{MSE}_\text{ext}(\E{\theta_\alpha[k]}{},\boldsymbol{\theta}_\text{NL}) =&\text{\space}\E{\vphantom{\left(\right)^2}{r}_{y_\alpha y_\alpha}(\boldsymbol{\theta}_\text{NL})-2{r}_{yy_\alpha}(\boldsymbol{\theta}_\text{NL})
				\left(\alpha_\text{d}+\E{\theta_{\alpha}[k]}{}\right)+{r}_{yy}(\boldsymbol{\theta}_\text{NL})\left(\alpha_\text{d}+\E{\theta_{\alpha}[k]}{}\right)^2}{}, 
		\end{split}
	\end{equation}
	where ${r}_{y_\alpha y_\alpha}(\boldsymbol{\theta}_\text{NL}) = \E{\left(y_{\alpha_\text{a}x}[k]+\mathbf{h}_{\alpha_\text{a}x}^T[k]\boldsymbol{\theta}_\text{NL}\right)^2}{}$. 
	Again with $\alpha_\text{d}$ separated from \eqref{eq:mmseSolutionAlpha1} and with the previously introduced error variable $v_\alpha[k]$, the equation above may be simplified to
	\begin{equation}
		\begin{split}
			J^\text{MSE}_\text{ext}(\E{\theta_\alpha[k]}{},\boldsymbol{\theta}_\text{NL}) =&\text{\space} {r}_{y_\alpha y_\alpha}(\boldsymbol{\theta}_\text{NL})- \frac{{r}_{yy_\alpha}^2(\boldsymbol{\theta}_\text{NL})}{{r}_{yy}(\boldsymbol{\theta}_\text{NL})}+{r}_{yy}(\boldsymbol{\theta}_\text{NL})\E{v_\alpha[k]}{}^2.
		\end{split}
	\end{equation}
	Finally, one can see that with \eqref{eq:convAlpha} and an appropriately chosen step-size
	\begin{equation}
		J^\text{MSE}_\text{ext}(\E{\theta_\alpha[k]}{},\boldsymbol{\theta}_\text{NL}) \leq J^\text{MSE}_\text{ext}(\E{\theta_\alpha[k-1]}{},\boldsymbol{\theta}_\text{NL}), 
		\label{eq:inequalpha}
	\end{equation}
	which means that the \ac{mse} does not increase in the mean for every update step of $\theta_\alpha[k]$, considering an arbitrary $\boldsymbol{\theta}_\text{NL}$.\\ Analogously, this analysis can be shown for the parameter vector $\boldsymbol{\theta}_\text{NL}[k]$. Therefore, the error vector $\mathbf{v}_\text{NL}[k] = \boldsymbol{\theta}_\text{NL}[k]-\boldsymbol{\theta}_{\text{NL},\text{W}}(\theta_{\alpha})$ is introduced with the \ac{xhec} Wiener solution $\boldsymbol{\theta}_{\text{NL},\text{W}}(\theta_\alpha)$ from (\ref{eq:mmseSolutionTheta1}), and an arbitrary but constant $\theta_\alpha[k] = \theta_\alpha$ is considered. Subtracting $\boldsymbol{\theta}_{\text{NL},\text{W}}(\theta_\alpha)$ from both sides of (\ref{eq:nlUpdate}), assuming that $\boldsymbol{\theta}_\text{NL}[k-1]$ is statistically independent of $\mathbf{h}_{\alpha_\text{a}x}[k]-\left(\alpha_\text{d}+\theta_\alpha[k]\right)\mathbf{h}_x[k]$, and computing the expected value yields
	\begin{equation}
		\begin{split}
			\E{\mathbf{v}_\text{NL}[k]}{} =&\text{\space} \E{\mathbf{v}_\text{NL}[k-1]}{}-\mu_\text{NL}\left(\mathbf{r}_{\mathbf{h}y}(\theta_{\alpha}) +\mathbf{R}_{\mathbf{hh}}(\theta_{\alpha})\E{\boldsymbol{\theta}_\text{NL}[k-1]}{}\right).
		\end{split}
		\label{eq:convNL1}
	\end{equation}
	Using $\mathbf{r}_{\mathbf{h}y}(\theta_{\alpha})$ from (\ref{eq:mmseSolutionTheta1}) allows rearranging (\ref{eq:convNL1}) as 
	\begin{equation}
		\E{\mathbf{v}_\text{NL}[k]}{} = \left(\mathbf{I}-\mu_\text{NL}\mathbf{R}_{\mathbf{hh}}(\theta_{\alpha})\right)\E{\mathbf{v}_\text{NL}[k-1]}{}, 
	\end{equation}
	with $\mathbf{I}$ being the identity matrix. It can be shown that the mean of the error $\E{\mathbf{v}_\text{NL}[k]}{}$ converges towards zero for $0<\mu_\text{NL}<2/\lambda_\text{max}(\mathbf{R}_{\mathbf{hh}}(\theta_{\alpha}))$, with $\lambda_\text{max}(\mathbf{R}_{\mathbf{hh}}(\theta_{\alpha}))$ as the maximum eigenvalue of $\mathbf{R}_{\mathbf{hh}}(\theta_{\alpha})$. Consequently, the parameter vector $\boldsymbol{\theta}_\text{NL}[k]$ converges in the mean towards the \ac{xhec} Wiener solution given $\theta_\alpha$. To interpret this behavior in terms of the \ac{mse} cost function, the constant $\theta_\alpha$ as well as $\E{\boldsymbol{\theta}_\text{NL}[k]}{}$ are inserted into \eqref{eq:mmseCostFunction2} such that 
	\begin{equation}
		\begin{split}
			J^\text{MSE}_\text{ext}(\theta_\alpha,\E{\boldsymbol{\theta}_\text{NL}[k]}{}) =&\text{\space} r_{\Delta y\Delta y}(\theta_\alpha) +2\mathbf{r}^T_{\mathbf{h}y}(\theta_\alpha)
			\E{\boldsymbol{\theta}_\text{NL}[k]}{}+\E{\boldsymbol{\theta}_\text{NL}^T[k]}{}\mathbf{R}_{\mathbf{hh}}(\theta_\alpha)\E{\boldsymbol{\theta}_\text{NL}[k]}{}
		\end{split}
		\label{eq:convLMSNL}
	\end{equation}
	with $r_{\Delta y\Delta y}(\theta_\alpha) = \E{\left(y_{\alpha x}[k]-\left(\alpha_\text{d}+\theta_{\alpha}\right)y_x[k]\right)^2}{}$. Replacing every instance of $\E{\boldsymbol{\theta}_\text{NL}[k]}{}$ in the equation above by $(\E{\boldsymbol{\theta}_\text{NL}[k]}{}-\mathbf{0})$ with the zero vector $\mathbf{0} = \boldsymbol{\theta}_{\text{NL},\text{W}}+\mathbf{R}_{\mathbf{hh}}^{-1}(\theta_\alpha)\mathbf{r}_{\mathbf{h}y}(\theta_{\alpha})$, obtained from \eqref{eq:mmseSolutionTheta1}, allows rewriting \eqref{eq:convLMSNL} as  
	\begin{equation}
		\begin{split}
			J^\text{MSE}_\text{ext}(\theta_\alpha,\E{\boldsymbol{\theta}_\text{NL}[k]}{}) =&\text{\space}  r_{\Delta y\Delta y}(\theta_\alpha) -\mathbf{r}^T_{\mathbf{h}y}\mathbf{R}_{\mathbf{hh}}^{-1}(\theta_\alpha)
			\mathbf{r}_{\mathbf{h}y}(\theta_{\alpha})+\E{\mathbf{v}_\text{NL}^T[k]}{}\mathbf{R}_{\mathbf{hh}}(\theta_\alpha)\E{\mathbf{v}_\text{NL}[k]}{}
		\end{split}
	\end{equation}
	Ultimately, one can see with \eqref{eq:convNL1} and a correctly chosen step-size $\mu_\text{NL}$ that
	\begin{equation}
		J^\text{MSE}_\text{ext}(\theta_\alpha,\E{\boldsymbol{\theta}_\text{NL}[k]}{})\leq J^\text{MSE}_\text{ext}(\theta_\alpha,\E{\boldsymbol{\theta}_\text{NL}[k-1]}{}). 
		\label{eq:ineqNL}
	\end{equation}
	Thus, the \ac{mse} does not increase in the mean for every update step of $\boldsymbol{\theta}_\text{NL}[k]$, considering an arbitrary $\theta_\alpha$. \\
	As the two update equations in the \ac{xhec} \ac{sgd} approach are executed alternatively, the separate investigations of the two update equations are combined in the following and graphically illustrated in Fig.~\ref{fig:nonConvexCost_lms}. Again, the black line represents the \ac{mse} cost function $J^\text{MSE}_\text{ext}(\theta_\alpha,\boldsymbol{\theta}_\text{NL,W}(\theta_\alpha))$, where $\boldsymbol{\theta}_\text{NL}$ in (\ref{eq:mmseCostFunction2}) is replaced by the Wiener solution from (\ref{eq:mmseSolutionTheta1}). Next, considering an initial $\boldsymbol{\theta}_\text{NL}[0]$ yields the temporary cost function $J^\text{MSE}_\text{ext}(\theta_\alpha,\boldsymbol{\theta}_\text{NL}[0])$, illustrated by the dashed dark gray line in Fig.~\ref{fig:nonConvexCost_lms}. In this cost function the \ac{xhec} Wiener filter directly estimates the minimum (black marker (\Romannum{1}$^\prime$)). The \ac{xhec} \ac{sgd} approach, however, only takes a step towards this \ac{xhec} Wiener solution in the mean (gray marker (\Romannum{1})) from the starting point (\Romannum{0}). In the second step the \ac{xhec} Wiener solution estimates the optimum $\boldsymbol{\theta}_\text{NL,W}(\theta_\alpha[1])$ given the previous \ac{sgd} estimate of $\theta_{\alpha}[1]$. In Fig.~\ref{fig:nonConvexCost_lms}, this is illustrated by the black marker (\Romannum{2}$^\prime$). The \ac{xhec} \ac{sgd} approach again takes in the mean a step towards this \ac{xhec} Wiener solution (gray marker  (\Romannum{2})). After this, the described procedure repeats, starting with the new temporary cost function $J^\text{MSE}_\text{ext}(\theta_\alpha,\boldsymbol{\theta}_\text{NL}[1])$, which is illustrated by the dashed gray line in Fig.~\ref{fig:nonConvexCost_lms}.\\Consequently, one can see with \eqref{eq:inequalpha} and \eqref{eq:ineqNL} that \[\textstyle J^\text{MSE}_\text{ext}(\E{\theta_\alpha[0]}{},\boldsymbol{\theta}_\text{NL}[0])\geq J^\text{MSE}_\text{ext}(\E{\theta_\alpha[1]}{},\boldsymbol{\theta}_\text{NL}[0]) \geq 
	J^\text{MSE}_\text{ext}(\E{\theta_\alpha[1]}{},\E{\boldsymbol{\theta}_{\text{NL}}[1]}{}) \geq J^\text{MSE}_\text{ext}(\E{\theta_\alpha[2]}{},\E{\boldsymbol{\theta}_\text{NL}[1]}{})\geq\ldots,\]
	thus, the \ac{xhec} \ac{sgd} update steps do not increase the costs in the mean. This procedure is exemplary illustrated by the gray markers (\Romannum{0})-(\Romannum{3}) in Fig.~\ref{fig:nonConvexCost_lms}. 
	\\ The identified stability bounds for $\mu_\text{NL}$ and $\mu_\alpha$, require statistical knowledge which might be unknown in practice. In order to determine a feasible step-size without requiring any statistical knowledge, the apriori error before an update step is compared to the posterior error after an update step. Therefore, the posterior error of $\theta_\alpha$ is written as 
	\begin{equation}
		\begin{split}
			\acute{e}_\alpha[k] =&\text{\space} y_{\alpha_\text{a} x}[k]+\mathbf{h}_{\alpha_\text{a} x}^T[k]\boldsymbol{\theta}_\text{NL}[k-1]-\left(\alpha_\text{d}+\theta_\alpha[k]\right)\left(y_x[k]+\mathbf
			{h}_x^T[k]\boldsymbol{\theta}_\text{NL}[k-1]\right).
		\end{split}
		\label{eq:alphaPosterior}
	\end{equation}
	By inserting (\ref{eq:alphaUpdate}) in (\ref{eq:alphaPosterior}), the posterior error may be written as a function of the apriori error from \eqref{eq:apriorierr_alpha} as
	\begin{equation}
		\acute{e}_\alpha[k] = \left(1-\mu_\alpha\left(y_x[k]+\mathbf
		{h}_x^T[k]\boldsymbol{\theta}_\text{NL}[k-1]\right)^2\right)e_\alpha[k].
	\end{equation}
	As can be seen from the equation above, with the step-size $0<\mu_\alpha\leq 2/\left(y_x[k]+\mathbf
	{h}_x^T[k]\boldsymbol{\theta}_\text{NL}[k-1]\right)^2$ the instantaneous squared error does not increase in one update step. Consequently, if this condition holds, for all $k$, \eqref{eq:alphaUpdate} does not diverge. To simplify this bound, the term $y_x[k]+\mathbf
	{h}_x^T[k]\boldsymbol{\theta}_\text{NL}[k-1]$ is approximated by the ideal \ac{adc} output. Since the maximum possible \ac{adc} output value $y_x^\text{max}$ is well defined by the number of bits, $2 / \left(y_x^\text{max}\right)^2$ represents a constant upper bound for the step-size. Analogously, a feasible step-size bound for $\mu_\text{NL}$ is derived with the posterior error of $\boldsymbol{\theta}_\text{NL}$ as
	\begin{equation}
		\begin{split}
			\acute{e}_\text{NL}[k] =&\text{\space} y_{\alpha_\text{a} x}[k]+\mathbf{h}_{\alpha_\text{a} x}^T[k]\boldsymbol{\theta}_\text{NL}[k]-\left(\alpha_\text{d}+\theta_\alpha[k]\right)\left(y_x[k]+\mathbf
			{h}_x^T[k]\boldsymbol{\theta}_\text{NL}[k]\right).
		\end{split}
		\label{eq:nlPosterior}
	\end{equation}
	By inserting the apriori error from \eqref{eq:apriorierr_nl} into \eqref{eq:nlPosterior}, the posterior error may be expressed as
	\begin{equation}
		\begin{split}
			\acute{e}_\text{NL}[k] =&\text{\space} \big(1-\mu_\text{NL}\left(\mathbf{h}_{\alpha_\text{a}x}[k]-\left(\alpha_\text{d}+\theta_\alpha[k]\right)\mathbf{h}_x[k]\right)^T
			\left(\mathbf{h}_{\alpha_\text{a}x}[k]-\left(\alpha_\text{d}+\theta_\alpha[k]\right)\mathbf{h}_x[k]\right)\big)e_\text{NL}[k].
		\end{split}
	\end{equation}
	The equation above shows that for $0<\mu_\text{NL}\leq2/||\mathbf{h}_{\alpha_\text{a}x}[k]-\left(\alpha_\text{d}+\theta_\alpha[k]\right)\mathbf{h}_x[k]||_2^2$ the instantaneous squared error does not increase within one update step. Considering practically meaningful values for the scaling factor mismatch allows neglecting  $\theta_\alpha[k]$, thus the step-size bounds 
	\begin{equation}
		0<\mu_\text{NL}\leq2 / \max\limits_{k}||\Delta\mathbf{h}[k]||_2^2
	\end{equation}
	from \cite{1_HEC} yield a good approximation.
	\section{RESULTS}
	\label{se:Results}
	In the following, the \ac{xhec} approach is analyzed via behavioral Matlab simulations, with respect to the impact of excluded non-idealities, analog input noise, scaling factor mismatch as well as performance dependency regarding the absolute value of the scaling factor. Additionally, measurements are carried out on 24 \acp{adc} integrated in state-of-the-art radar sensors. For these measurements only already available on-chip test equipment is utilized, which verifies the low hardware requirements of the proposed calibration technique.
	\subsection{Simulation Results}
\begin{figure}[t]
		\centering
		\includegraphics{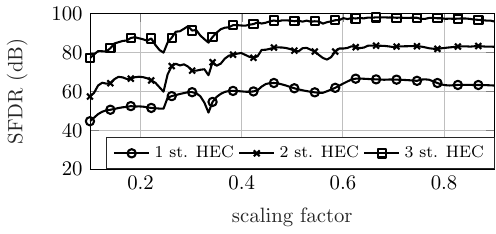}	
		\caption{Simulated \ac{sfdr} performance over a scaling factor-sweep. The illustrated results are averaged over 100 \acp{adc}.}
		\label{fig:sfdr_vs_scaling}
	\end{figure}

\begin{figure}[t]
		\centering
		\begin{minipage}{0.49\linewidth}
			\centering\includegraphics{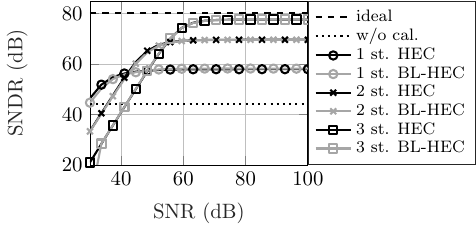}\\
			\centering (a)
		\end{minipage}\hfill
		\begin{minipage}{0.49\linewidth}
			\centering\includegraphics{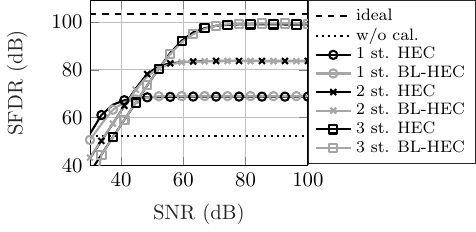}\\
			\centering (b)
		\end{minipage}
		\caption{Simulated (a) \ac{sndr} and (b) \ac{sfdr} performance over \ac{snr}. The illustrated results are averaged over 100 \acp{adc}.}
		\label{fig:sndr_sfdrvsnoise}
	\end{figure}

	The simulations model a 13 bit pipelined \ac{adc} with six stages and a sampling rate of $100\,\text{MHz}$. The first five stages are modeled with $2.5\,\text{bits}$ and ideal stage gains $G_i = 4$, while the final flash \ac{adc} stage includes $3\,\text{bits}$. The stages 1-5 incorporate gain and \ac{dac} mismatches. These mismatches are chosen from a uniform \ac{pdf}, such that $\zeta_i \sim \mathcal{U}\left[-25\,\text{LSB},25\,\text{LSB}\right]$ and $e_{\text{s},i}^\text{DA}\sim\mathcal{U}\left[-15\,\text{LSB},15\,\text{LSB}\right]$. Therefore, gain and \ac{dac} mismachtes in the $i$th stage contribute to the overall \ac{adc} nonlinearity with a maximum value of $25/\left(4^{i-1}\right)\,\text{LSB}$ and $15/\left(4^{i-1}\right)\,\text{LSB}$, respectively. If not denoted otherwise, the first three stages are covered by the post-correction model. A full-scale sinusoid\footnote{Note that, in practice an input signal slightly below full-scale ($-1\ldots-2\,\text{dBFS}$) is used. As long as all output levels of the first stage are covered by the test signal, this would not limit the calibration performance of the algorithm.} at $10.77\,\text{MHz}$ is used as the input signal, and the scaling factor mismatch is taken from a normal distribution $\delta \sim \mathcal{N}\left(0,10^{-4}\right)$ with zero mean and a variance of $10^{-4}$. For performance evaluations, the \ac{sndr} as well as the \ac{sfdr} are used, whereby the \ac{sfdr} is considered as the difference of the highest wanted peak and the highest unwanted peak in the spectrum. If results are shown based on the \ac{hec} Wiener filter or the \ac{xhec} Wiener filter, all required statistics are estimated via the corresponding sample means using $2\cdot 10^3$ samples.
	\paragraph*{Choice of Scaling Factor}In order to determine a valid scaling factor, the \ac{sfdr} performance of the \ac{hec} Wiener filter is evaluated over a scaling factor-sweep of $\alpha_{\text{a}} = \alpha_{\text{d}}$ from $0.1$ to $0.9$. Fig.~\ref{fig:sfdr_vs_scaling} shows the \ac{sfdr} averaged over 100 simulated \acp{adc} with one, two, and three stages considered for calibration. As can be seen, the achieved \ac{sfdr} decreases for scaling factors smaller than $0.5$. However, as for scaling factors greater than $0.5$, the \ac{hec} approach performs well, a scaling factor of $\frac{1}{\sqrt{2}}$ is used for all following investigations. 
	\paragraph*{Excluded Non-Idealities}To investigate the impact of non-ideal excluded stages, derived in \eqref{eq:mmseSolution2}, the \ac{hec} Wiener filter from \cite{1_HEC} is used for calibration. The post-correction model includes the first, the first two and the first three stages. All included stages are simulated as ideal, i.e., $\boldsymbol{\theta}_0 = \mathbf{0}$, such that only the impact of excluded stages is present. Furthermore, the simulations are performed with a sufficiently high \ac{snr} as well as $\delta = 0$. Table\,\ref{tab:xNonlin} shows that the \ac{hec} approach improves both \ac{sndr} and \ac{sfdr} slightly although the non-ideal stages are not included in the post-correction model.  
	\begin{table}[t!]
		\centering
		\caption{Effect of non-ideal excluded stages based on simulated \ac{sndr} and \ac{sfdr} performance with respect to 100 \acp{adc}.}
		\begin{tabular}{|l|l||c|c|c|c|c|c|c|}
			\cline{3-8}
			\multicolumn{2}{c||}{\multirow{2}{*}{}}	& \multicolumn{3}{c|}{SNDR (dB)} & \multicolumn{3}{c|}{SFDR (dB)} 			\\\cline{3-8}
			\multicolumn{2}{c||}{}			 	& min 	& mean & max	 		    & min 	& mean 	& max		\\\hline\hline
			\multirow{2}{*}{1 stage} &w/o cal.	&52.7	&56.5  &62.1			    &58.6	&67.0	&74.8		\\\cline{2-8}	
			&w cal.    &51.3	&57.6  &68.4  				&60.8	&68.3	&81.7 		\\\hline				  
			\multirow{2}{*}{2 stages}&w/o cal.  &64.0	&68.1  &72.2				&75.5	&82.8	&89.5		\\\cline{2-8}	
			&w cal.	&64.2   &69.9  &75.5  				&75.8	&84.2	&93.7		\\\hline
			\multirow{2}{*}{3 stages}&w/o cal.  &75.0	&76.8  &78.6				&92.4	&97.7	&103.6		\\\cline{2-8}
			&	w cal.	&75.1	&77.4  &79.2				&91.9	&98.3	&103.6		\\\hline							  
		\end{tabular}
		\label{tab:xNonlin}
	\end{table}
	\paragraph*{Analog Input Noise}The impact of analog noise on the \ac{sndr} and \ac{sfdr} performance is analyzed in Fig.~\ref{fig:sndr_sfdrvsnoise}. Therefore, the performance of the \ac{hec} Wiener filter under ideal conditions, i.e., $\delta = 0$, and the \ac{xhec} Wiener filter, with $\delta \sim \mathcal{N}\left(0,10^{-4}\right)$, is evaluated for 100 different \acp{adc} considering an \ac{snr}-sweep from $30\,\text{dB}$ to $100\,\text{dB}$. Each \ac{adc} is calibrated with one, two and three stages included in the post-correction model. In Fig.~\ref{fig:sndr_sfdrvsnoise}\,(a), one can see that the \ac{xhec} approach achieves comparable \ac{sndr} results as the \ac{hec} approach under ideal conditions. As can be seen, calibrating more stages requires higher \ac{snr} to maximize the \ac{sndr} results.
	A similar behavior may be observed in Fig.~\ref{fig:sndr_sfdrvsnoise}\,(b) for the achieved \ac{sfdr}. Based on these results, the following simulations are carried out with an \ac{snr} of 70\,dB.
	
\begin{figure}[t]
		\centering
		\begin{minipage}{0.49\linewidth}
			\centering\includegraphics{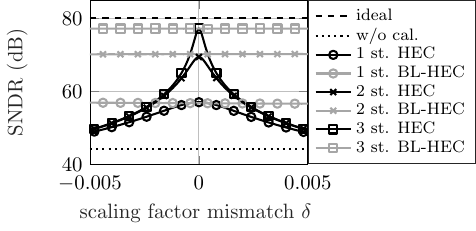}\\
			\centering (a)
		\end{minipage}\hfill
		\begin{minipage}{0.49\linewidth}
			\centering\includegraphics{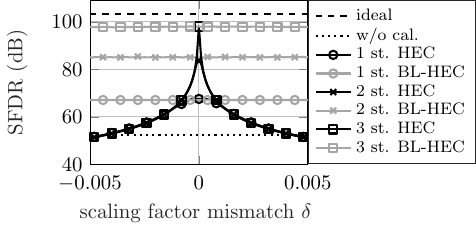}\\
			\centering (b)
		\end{minipage}
		\caption{Simulated (a) \ac{sndr} and (b) \ac{sfdr} performance over scaling factor mismatch. The illustrated results are averaged over 100 \acp{adc}.}
		\label{fig:sndr_sfdrvsdelta}
	\end{figure}

\begin{figure}[t]
		\centering
		\begin{minipage}{0.49\linewidth}
			\centering\includegraphics{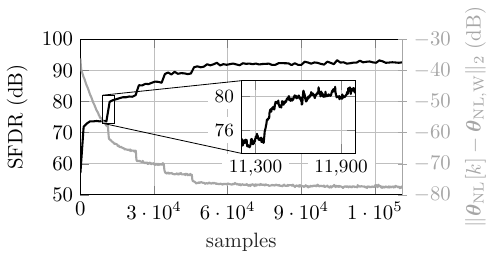}\\
			\centering (a)
		\end{minipage}\hfill
		\begin{minipage}{0.49\linewidth}
			\centering\includegraphics{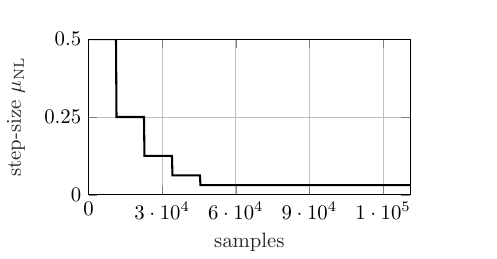}\\
			\centering (b)
		\end{minipage}
		\caption{(a) Simulated \ac{sfdr} and error norm over number of samples used for calibration. The illustrated results are averaged over 100 \acp{adc}. (b) Step-size reduction of $\mu_\text{NL}$.}
		\label{fig:sndr_sfdrvsdeltaxhex}
	\end{figure}

	\paragraph*{Scaling Factor Mismatch}
	To analyze the impact of mismatching scaling factors, the \ac{hec} Wiener filter and the \ac{xhec} Wiener filter are simulated with different numbers of stages included in the post-correction model and a scaling factor mismatch $\delta$ from $-5\cdot10^{-3}$ to $5\cdot10^{-3}$. As can be seen in Fig.~\ref{fig:sndr_sfdrvsdelta}, already small scaling factor mismatches cause a significant performance degradation in terms of \ac{sndr} and \ac{sfdr} for the \ac{hec} approach. However, the proposed \ac{xhec} approach achieves a constant \ac{sndr} and \ac{sfdr}, even for mismatching scaling factors. Note that, the applied scaling factor mismatch is chosen according to a practically relevant range identified via measurements. For larger scaling factor mismatches, the performance of the \ac{xhec} approach would show a scaling factor dependency.
	\paragraph*{\ac{xhec} \ac{sgd} Approach}
	In order to achieve a fast convergence and a small steady state error with the \ac{xhec} \ac{sgd} approach, the step-size $\mu_\text{NL}$ is chosen to be variable according to the pattern illustrated in Fig.~\ref{fig:sndr_sfdrvsdeltaxhex}\,(b). The step-size $\mu_\alpha$ was chosen as $\mu_\alpha = 0.5\cdot\mu_\text{NL}$ and, therefore, also reduces over time. While step-size reduction is a well known procedure with many existing reduction strategies \cite{mandic2009complex}, the one applied within this work is found empirically. Note that, all used step-sizes are a power of two, which streamlines a hardware implementation. Fig.~\ref{fig:sndr_sfdrvsdeltaxhex}\,(a) shows the achieved \ac{sfdr}  and the error norm $||\boldsymbol{\theta}_\text{NL}[k]-\boldsymbol{\theta}_{\text{NL},\text{W}}||_2$, with both lines being averaged over 100 simulated \acp{adc}. As can be seen, the \ac{xhec} \ac{sgd} approach achieves on average $91.85\,\text{dB}$ \ac{sfdr} after $4.8\cdot 10^4$ samples. In comparison to the \ac{hec} approach from \cite{1_HEC}, the \ac{xhec} approach shows comparable calibration results even for mismatching scaling factors. However, the necessary number of samples increases from $3\cdot 10^4$ to $4.8\cdot 10^4$. A comprehensive summary of the simulation results, including a comparison with other calibration techniques, is presented in Table\,\ref{tab:comp}.
	\subsection{Measurement Results}
\begin{figure}[t!]
		\centering
		\includegraphics{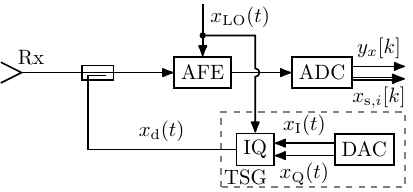}
		\caption{Block diagram of the radar receiver including the internal \ac{tsg}.}
		\label{fig:blockdiagmeas}
	\end{figure}

\begin{figure}[t]
		\centering
		\includegraphics{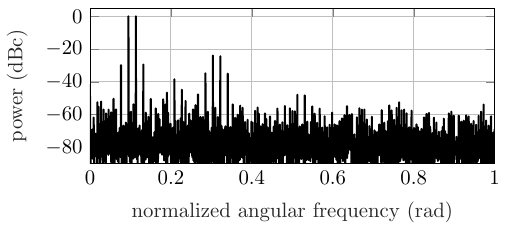}	
		\caption{Spectrum of the on-chip generated test signal used for the \ac{adc} calibration.}
		\label{fig:TestSignal}
	\end{figure}

\begin{figure}[t]
		\centering
		\includegraphics{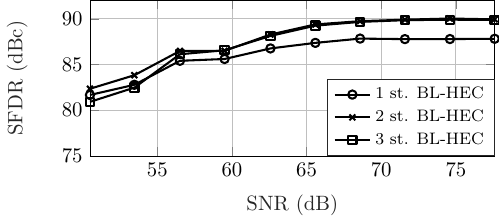}	
		\caption{Measured \ac{sfdr} performance over \ac{snr}. The illustrated results are averaged over 24 \acp{adc}.}
		\label{fig:sndr_sfdrvssnr_meas}
	\end{figure}

\begin{figure}[t]
		\centering
		\includegraphics{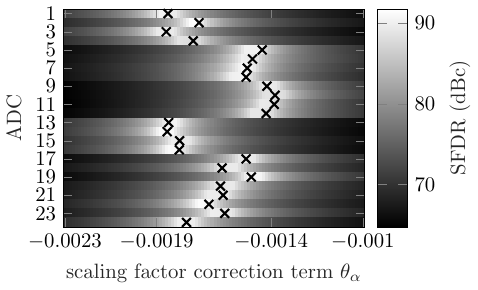}	
		\caption{Measured \ac{sfdr} performance over a scaling factor correction term.} 
		\label{fig:sfdrvsdelta_meas}
	\end{figure}

\begin{figure}[t]
		\centering
		\includegraphics{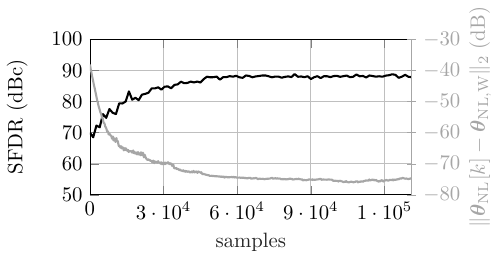}
		\caption{Measured \ac{sfdr} and error norm over number of samples used for calibration. The illustrated results are averaged over 24 \acp{adc}.}
		\label{fig:SFDRvsSamples_lms_meas}
	\end{figure}

	To showcase the low requirements on the analog hardware of the proposed algorithm, the measurements are carried out on 24 \acp{adc} integrated in state-of-the-art $77\,\text{GHz}$ radar sensors. Therein, the \ac{adc}, employed in the base-band of the receive path, should be calibrated by only using internally available test signals. Fig.~\ref{fig:blockdiagmeas} shows a high-level block diagram of the receive path with the \ac{afe}, i.e., a mixer, a low-noise amplifier, and several filters, followed by a pipelined \ac{adc}. Furthermore, the radar sensor is equipped with a \ac{tsg} for standard monitoring tasks, such as gain and phase monitoring \cite{wagner2023accurate,fujibayashi_76-_2017}. This enables injecting a test signal with an amplitude 85\% of the full-scale into the receive path directly after the receive antenna. The \ac{tsg} provides a two-tone test signal, which is generated by a low resolution \ac{dac} and afterwards modulated onto a carrier signal. Therefore, the spectrum of the two-tone test signal, illustrated in Fig.~\ref{fig:TestSignal}, shows a substantial amount of quantization noise originating from the low \ac{dac} resolution. Furthermore, the test signal contains strong second, third, and fifth order intermodulation products and harmonics. Consequently, the low spectral purity of the test signal prevents a straight forward \ac{adc} calibration.\\  To use this signal for the \ac{xhec} approach, in addition to the test signal, a scaled version of it has to be fed into the \ac{adc}. This is achieved by changing the receiver gain in the \ac{afe} by $-3\,\text{dB}$ for the second measurement. Therefore, a scaling of $\frac{1}{\sqrt{2}}$ can be performed without any changes of the existing hardware. Furthermore, the scaled and the unscaled \ac{adc} raw data, including the stage outputs, are recorded, and then further processed using Matlab\footnote{Note that, further processing involves limiting the \ac{adc} to 13 bit for comparison.}.\\
	Comparing the described measurement setup from Fig.~\ref{fig:blockdiagmeas} with the \ac{hec} approach block diagram in Fig.~\ref{fig:hec}, shows that the measurements are conducted under adverse conditions. As the test signal is injected directly after the receive antenna, it propagates through the whole receive path including amplification, downconversion, and filtering before it reaches the \ac{adc}. Along with the test signal, also noise from the antenna is amplified, resulting in a lower \ac{snr} compared to an ideal implementation, where the test signal is injected directly into the \ac{adc} and receiver gain is reduced. To cope with this issue, several periods of the test signal are averaged in post-processing until a desired \ac{snr} is reached.\\
	After estimating the calibration parameters with the proposed methods, the performance is evaluated by the achieved \ac{sfdr}. Therefore, a spectrally pure two-tone test signal from an external test signal source is injected into the receiver while the \ac{adc} raw data is recorded. Next, the previously obtained calibration parameters are applied and the \ac{sfdr} values are computed, by means of taking the difference of the highest wanted peak and the highest unwanted peak in the spectrum. 
	The measurable \ac{sndr} is limited by the analog noise at the \ac{adc} input due to the measurement setup, and is therefore not stated.\\
	If not denoted otherwise, the internally as well as the externally generated two-tone test signal is employed with the normalized angular frequencies $2\pi\frac{f_1}{f_\text{s}} =0.0942\,\text{rad}$ and $2\pi\frac{f_2}{f_\text{s}} =0.11\,\text{rad}$, with $f_1$ and $f_2$ as the tone frequencies and $f_\text{s}$ as the sampling frequency. Furthermore, all statistics required for the \ac{hec} Wiener filter and the \ac{xhec} Wiener filter are estimated via the corresponding sample mean using $2\cdot10^3$ samples. 
	\paragraph*{Analog Input Noise}
	Fig.~\ref{fig:sndr_sfdrvssnr_meas} shows the \ac{sfdr} performance of the \ac{xhec} Wiener filter, averaged over 24 \acp{adc}, with respect to the \ac{snr} for different numbers of stages included in the post-correction model. As mentioned above, the different \ac{snr} values are obtained by averaging over multiple periods of the test signal, thus, this represents the \ac{snr} at the \ac{adc} output. One can see that, as with the simulation results in Fig.~\ref{fig:sndr_sfdrvsnoise}\,(b), an \ac{snr} of approximately $70\,\text{dB}$ is required to obtain maximum calibration performance. Furthermore, adding a third stage to the post-correction model does not improve the performance, which indicates less gain and \ac{dac} errors than assumed in the simulations. Consequently, in the following, only the first two stages are considered for calibration while an \ac{snr} of $70\,\text{dB}$ is utilized.  
	\paragraph*{Scaling Factor Mismatch}
	To investigate the impact of a scaling factor mismatch, the \ac{sfdr} performance of the \ac{hec} Wiener filter from \cite{1_HEC} is evaluated, whereby the digital scaling factor is set to $\frac{1}{\sqrt{2}}$ and a scaling factor correction term $\theta_{\alpha}$ is swept from $-2.3\cdot10^{-3}$ to $-1\cdot10^{-3}$, thus, $\alpha_{\text{d}} = \frac{1}{\sqrt{2}}+\theta_{\alpha}$. This procedure creates an \ac{sfdr} curve similar to Fig.~\ref{fig:sndr_sfdrvsdelta}\,(b), however, with the maximum \ac{sfdr} centered around the scaling factor correction term which compensates the scaling factor mismatch. The results are illustrated in Fig.~\ref{fig:sfdrvsdelta_meas}, where each row corresponds to the \ac{sfdr} of a single \ac{adc}. Extrapolating Fig.~\ref{fig:sfdrvsdelta_meas} on the right side to $\theta_{\alpha} = 0$ corresponds to the performance of the original \ac{hec} approach and would obviously result in a poor \ac{sfdr}. The results of the proposed \ac{xhec} Wiener filter, with $\alpha_\text{d} = \frac{1}{\sqrt{2}}$ and the estimated scaling factor correction terms, are illustrated by the black crosses in Fig.~\ref{fig:sfdrvsdelta_meas}, which are located close to the maximum \ac{sfdr} values for all 24 \acp{adc}. These results highlight the practical relevance of the \ac{xhec} approach proposed in this work.
	\paragraph*{\ac{xhec} \ac{sgd} Approach}
	Although the \ac{xhec} Wiener filter performs well in terms of calibration performance, its computational complexity does not favor an on-chip implementation. Therefore, this work also introduces the adaptive \ac{xhec} \ac{sgd} approach. To obtain fast convergence as well as a small steady state error, the same step-size reduction as within the simulations is utilized for the measurements. Fig.~\ref{fig:SFDRvsSamples_lms_meas} shows the achieved \ac{sfdr} as well as the error norm $||\boldsymbol{\theta}_\text{NL}[k]-\boldsymbol{\theta}_{\text{NL},\text{W}}||_2$ averaged over 24 measured \acp{adc}. From the \ac{sfdr} curve, one can see that after $4.8\cdot 10^4$ samples the performance increase saturates, thus this number of samples is used for all following results of the \ac{xhec} \ac{sgd} approach. 
	\paragraph*{Frequency Dependency of the Calibration Performance}
	As \ac{sfdr} is related to the test signal frequency, this frequency dependency is analyzed in  Table\,\ref{tab:sfdr-performance}. Different sets of tone frequencies of the on-chip two-tone test signal are used to estimate the calibration parameters. The used normalized angular frequency sets are $\Omega_1 = \left\{0.0197, 0.0339\right\}\,\text{rad}$, $\Omega_2 = \left\{0.0949, 0.1137\right\}\,\text{rad}$, and $\Omega_3 = \left\{0.3160, 0.3280\right\}\,\text{rad}$. These sets are shown in the first row of Table\,\ref{tab:sfdr-performance}. Next, the \acp{adc} are calibrated with the previously obtained calibration parameters, and the \ac{sfdr} is computed from multiple reference measurements with different two-tone frequencies from $\Omega_1$, $\Omega_2$, and $\Omega_3$. Table\,\ref{tab:sfdr-performance} summarizes the \ac{sfdr} performances of all frequency combinations for the \ac{xhec} Wiener filter and the \ac{xhec} \ac{sgd} approach. As can be seen, the \ac{xhec} Wiener filter achieves \ac{sfdr} values larger than $84\,\text{dB}$ except when the parameters are estimated with frequencies from $\Omega_3$ and evaluated with frequencies from $\Omega_1$. The results from the \ac{xhec} \ac{sgd} approach confirm this behavior. Additionally, one can see that using frequencies from $\Omega_1$ for estimation yields the worst \ac{sfdr} results, which is because in this case, the \ac{xhec} \ac{sgd} approach is not fully converged after $4.8\cdot10^4$ samples. 
	\begin{table}[t]
		\centering
		\caption{SFDR Performance of \ac{xhec} \ac{sgd} and Wiener Filter averaged over 24 \acp{adc}.}
		\label{tab:sfdr-performance}
		\begin{tabular}{|l||l|c|c|c|}
			\hline
			est. freq. & eval. freq. & $\Omega_1$ & $\Omega_2$ & $\Omega_3$ \\ \hline\hline
			\multirow{2}{*}{$\Omega_1$} & \ac{sgd} (dBc)& 80.69 & 83.41 & 82.93 \\ \cline{2-5}
			&  Wiener (dBc)  & 86.48 & 89.44 & 88.05 \\ \hline
			\multirow{2}{*}{$\Omega_2$} &  \ac{sgd} (dBc) & 84.15 & 87.94 & 86.60 \\ \cline{2-5}
			&  Wiener (dBc)& 84.81 & 89.36 & 87.83 \\ \hline
			\multirow{2}{*}{$\Omega_3$} &  \ac{sgd} (dBc) & 79.77 & 84.99 & 85.44 \\ \cline{2-5}
			&  Wiener (dBc)& 79.98 & 86.46 & 87.13 \\ \hline
		\end{tabular}
	\end{table}
	\paragraph*{Comparison}
	Finally, the simulation as well as the measurement results of the proposed methods are compared with prior work \cite{4_ADC_DigitalBackgroundCalibrationWithHistogramOfDecisionPointsInPipelinedADCs,5_ADC_ADigitalBackgroundCalibrationSchemeForPipelinedADCsUsingMultipleCorrelationEstimation,7_ADC_StatisticsBasedDigitalBackgroundCalibrationOfResidueAmplifierNonlinearityInPipelinedADCs,2_ADC_EqualizationBasedDigitalBackgroundCalibrationTechniqueForPipelinedADCs,1_HEC} in Table\,\ref{tab:comp}. As can be seen, even though mismatching scaling factors are assumed for the simulations in this work, and inherently present in the measurements, the \ac{xhec} approach achieves similar high calibration performance as the original \ac{hec} approach in \cite{1_HEC}. However, with $4.8\cdot 10^3$ the \ac{xhec} approach requires about $60\,\text{\%}$ more samples for convergence than its original version. Nevertheless, it still shows a shorter convergence time than the methods in \cite{7_ADC_StatisticsBasedDigitalBackgroundCalibrationOfResidueAmplifierNonlinearityInPipelinedADCs,5_ADC_ADigitalBackgroundCalibrationSchemeForPipelinedADCsUsingMultipleCorrelationEstimation,4_ADC_DigitalBackgroundCalibrationWithHistogramOfDecisionPointsInPipelinedADCs,2_ADC_EqualizationBasedDigitalBackgroundCalibrationTechniqueForPipelinedADCs}. Furthermore, the achieved \ac{sfdr} improvement observed in the measurements is lower than those reported in \cite{7_ADC_StatisticsBasedDigitalBackgroundCalibrationOfResidueAmplifierNonlinearityInPipelinedADCs,4_ADC_DigitalBackgroundCalibrationWithHistogramOfDecisionPointsInPipelinedADCs,2_ADC_EqualizationBasedDigitalBackgroundCalibrationTechniqueForPipelinedADCs,1_HEC}. However when considering the absolute \ac{sfdr} after calibration, the proposed method achieves higher values than those stated in \cite{7_ADC_StatisticsBasedDigitalBackgroundCalibrationOfResidueAmplifierNonlinearityInPipelinedADCs,4_ADC_DigitalBackgroundCalibrationWithHistogramOfDecisionPointsInPipelinedADCs,2_ADC_EqualizationBasedDigitalBackgroundCalibrationTechniqueForPipelinedADCs}. This suggests that the \acp{adc} used for measurement already showed relatively high \ac{sfdr} performance before calibration, leaving only limited headroom for further improvement.    
	\begin{table*}[t]
		\begin{center}
			\caption{Comparison with other calibration techniques.}
			\label{tab:comp}
			\begin{tabular}{|l||c|c|c|c|c||c|c|c|c|}
				\hline
				\multirow{3}{*}{Ref.} & \multirow{3}{*}{\cite{4_ADC_DigitalBackgroundCalibrationWithHistogramOfDecisionPointsInPipelinedADCs}} & \multirow{3}{*}{\cite{5_ADC_ADigitalBackgroundCalibrationSchemeForPipelinedADCsUsingMultipleCorrelationEstimation}}& \multirow{3}{*}{\cite{7_ADC_StatisticsBasedDigitalBackgroundCalibrationOfResidueAmplifierNonlinearityInPipelinedADCs}} & \multirow{3}{*}{\cite{2_ADC_EqualizationBasedDigitalBackgroundCalibrationTechniqueForPipelinedADCs}} &\multirow{3}{*}{\cite{1_HEC}} & \multicolumn{4}{c|}{This Work}\\\cline{7-10}
				&&&&&& \multicolumn{2}{c|}{sim.}  & \multicolumn{2}{c|}{meas.}  \\\cline{7-8}\cline{8-10}
				&&&&&& \ac{sgd} & Wiener & \ac{sgd} & Wiener\\\hline\hline
				Additional analog circuit			&none				&none			&moderate		&low			&low			&low				&low			&low/none$^*$				&low/none$^*$			\\\hline
				Additional digital hardware			&low				&low			&low			&low			&low			&low				&high			&low				&high			\\\hline
				Samples for cal.   					&$1\cdot10^6$		&$1\cdot10^8$	&$5\cdot10^6$	&$5\cdot10^4$	&$3\cdot10^4$	&$4.8\cdot10^4$		&$2\cdot10^3$	&$4.8\cdot10^4$		&$2\cdot10^3$\\\hline
				SFDR (dB)							&75.8				&-				&82.3			&83				&95.5			&91.85				&94.23			&87.94$^\dag$				&89.36$^\dag$\\\hline
				SFDR impr. (dB)   	 				&40.8				&-				&25.5			&41				&43				&42					&44.57			&24					&26.28\\\hline
				SNDR (dB) 							&68.2				&73.4			&70.9			&68				&78.7			&76.1				&76.68			&-					&-\\\hline
				SNDR impr. (dB) 					&34.1				&27				&21				&28				&28				&32.6				&33.18			&-					&-\\\hline
				Resolution (bits)     					&12					&12				&12				&12				&13				&13					&13				&13					&13\\\hline
				\multicolumn{9}{l}{\footnotesize $^*$None, as the required option for scaling is already implemented in the utilized radar sensor.} \\
				\multicolumn{9}{l}{\footnotesize $^\dag$The \ac{sfdr} of the measurement results is obtained in dBc.} \\
			\end{tabular}
		\end{center}
	\end{table*}
	\section{CONCLUSION}
	\label{se:Con}
	
	In this work, different limitations of the \ac{hec} approach were investigated. Therewith, it was shown that the performance of the \ac{hec} approach is highly sensitive to a mismatch between the analog and the digital scaling factor. To cope with this issue, the \ac{xhec} Wiener filter and the \ac{xhec} \ac{sgd} approach were introduced. Besides a comprehensive stability analysis, the performance of the proposed \ac{xhec} approach was verified via behavioral Matlab simulations and measurements. The measurements were carried out on 24 integrated \acp{adc} and confirm high \ac{sfdr} values, low hardware requirements, as well as low number of samples needed for calibration even for mismatching scaling factors. Ultimately, the results were compared to prior work.
	
	\section{ACKNOWLEDGMENT}
	This work has been supported by the ”LCM - K2 Center for Symbiotic Mechatronics” within the framework of the Austrian COMET - K2 program.


\end{document}